\documentclass[twocolumn,secnumarabic,amssymb, nobibnotes, aps, prd]{revtex4-2}
\usepackage{natbib}
\usepackage{balance}
\usepackage{amsmath}
\usepackage{amsfonts}
\usepackage{amssymb}
\usepackage{makeidx}
\usepackage{graphicx}
\usepackage{blindtext}

\begin{document}

\title{Bimodular Continuous Attractor Neural Networks with Static and Moving Stimuli}%

\author{Min Yan$^{1,*}$, Wen-Hao Zhang$^{2, 3}$, He Wang$^{1, 4}$, and K. Y. Michael Wong$^{1}$}%
\affiliation{$^1$Department of Physics, Hong Kong University of Science and Technology, Hong Kong SAR, P. R. China \\ 
$^2$Lyda Hill Department of Bioinformatics, UT Southwestern Medical Center, Dallas Texas, USA \\
$^3$O’Donnell Brain Institute, UT Southwestern Medical Center, Dallas  Texas, USA \\
$^4$Hong Kong University of Science and Technology Shenzhen Research Institute, Shenzhen 518057, China } 
\date{\today}%
\begin{abstract}
We investigated the dynamical behaviors of bimodular CANNs, each processing a modality of sensory input and interacting with each other. We found that when bumps coexist in both modules, the position of each bump is shifted towards the other input when the inter-modular couplings are excitatory and is shifted away when inhibitory.  When one inter-modular coupling is excitatory while another is moderately inhibitory, temporally modulated population spikes can be generated. On further increase of the inhibitory coupling, momentary spikes will emerge.  In the regime of bump coexistence, bump heights are primarily strengthened by excitatory inter-modular couplings but there is a lesser weakening effect due to a bump being displaced from the direct input.  When bimodular networks serve as decoders of multisensory integration, we extend the Bayesian framework to show that excitatory and inhibitory couplings encode attractive and repulsive priors respectively. At low disparity, the bump positions decode the posterior means in the Bayesian framework, whereas at high disparity, multiple steady states exist. In the regime of multiple steady states, the less stable state can be accessed if the input causing the more stable state arrives after a sufficiently long delay. When one input is moving, the bump in the corresponding module is pinned when the moving stimulus is weak, unpinned at intermediate stimulus strength, and tracks the input at strong stimulus strength, and the stimulus strengths for these transitions increase with the velocity of the moving stimulus. These results are important to understanding multisensory integration of static and dynamic stimuli.

\end{abstract}
\maketitle

\section{Introduction}
The human brain is sophisticated and advanced. It performs computations efficiently\cite{amit1992modeling,gerstner2014neuronal,dayan2003theoretical}. The brain receives inputs from surrounding environment all the time via different sensory modalities, e.g., visual, auditory, olfactory and vestibular, and so on. Experiments showed that different cortical regions in the brain are not completely isolated from each other, and there exist interactions between  different sensory modalities. When the brain is processing information, it is able to combine  cues coming from different sensory modalities, producing responses with higher accuracy or speed.  In addition,  this kind of multisensory integration can also give rise to some interesting behaviors, such as sensory illusion and response enhancement. 

Various models have been built to elucidate the information processing mechanism of the brain  \cite{ben1997traveling, wilson1972excitatory,amari1977dynamics,deneve1999reading,samsonovich1997path,camperi1998model}.  In this paper we study the bimodular version of a model that has gained widespread attention known as the the Continuous attractor neural  network (CANN) \cite{wu2002population,wu2005computing,wu2008dynamics,fung2008dynamics,fung2010moving,fung2012dynamical,fung2013resolution}. Due to their property of translational invariance of neuronal activities, these networks are endowed with the ability to hold a continuous family of stationary states \cite{camperi1998model,wu2002population,wu2005computing,wu2008dynamics}.  This feature enables the network to track a moving stimulus continuously, providing a convincing model of processing continuous information in the brain~\cite{wimmer2014bump,kim2008benefits,green2017neural,burak2009accurate}.

The CANNs have been studied extensively \cite{xie2002double,latham2003optimal, boucheny2005continuous, fung2008dynamics, burak2009accurate, fung2010moving, fung2012dynamical, fung2013resolution, wang2015rich, burak2012fundamental, seeholzer2019stability}. The ability to hold a continuous family of stationary states endows the network with the capacity to simulate and study different functions in the brain. For example, the working memory is crucial during information transmission and processing, and single-module CANNs can be used to illustrate the dynamics and decay of the working memory in the brain since the network can retain the memory over a longer period compared with those in single network constituents \cite{seung1996brain, seeholzer2019stability, burak2012fundamental}.  Besides the study of working memory, the CANN is also exploited in the investigations of path integration \cite{samsonovich1997path, burak2009accurate}. Scientists successfully generated grid-cell-like responses similar to those observed experimentally in the rat's brain, and suggested the CANN as the underlying mechanism responsible for the emergence of grid cells and for path integration. Another wide application of the CANN is in the explorations of the “head direction” \cite{xie2002double, boucheny2005continuous}. By inserting different neuron modules (populations), scientists realized the approximations of the head-velocity independent tuning curves observed in the postsubiculum (POS) and anticipatory responses observed in the anterior dorsal thalamus (ADT) \cite{xie2002double}. Besides, the directional representation encoded by the attractor network can be rapidly updated by external cues, which is consistent with experimental measurements in thalamic head direction cells \cite{boucheny2005continuous}.  

In terms of network architecture, researchers modified the architecture of the CANN according to experiments, including the properties of couplings and the number of modules or neuron populations. By building neural networks that are more biologically plausible, experiments and biological mechanisms can be explained convincingly. In one-dimensional CANNs, scientists introduced more than one dynamical variables to simulate the synaptic plasticity (short-term facilitation and short-term depression), which are able to illustrate memory retention and mobility improvement that gives rise to anticipative responses \cite{seeholzer2019stability, fung2013resolution, wang2015rich}. When generalizing the CANN to a two-dimensional structure, it is endowed with more abundant dynamical patterns \cite{fung2015spontaneous}, which means it has the potential to simulate and explain more complicated mechanisms, such as illustrating the responses of grid cells in implementing accurate path integration tasks \cite{burak2009accurate}. In addition, multi-dimensional attractor networks are also shown to perform reliable and optimal computations \cite{latham2003optimal} with noisy population codes, explaining that the brain can still perform optimal computations even when the reliabilities of cues are varying. 

One outstanding advantage of the CANN is the ability to perform optimal computations especially when dealing with noisy input signals. Scientists have explored that the multi-dimensional CANN can still compute reliably (reaching the Cram\'{e}r-Rao bound) for cues with varying reliability in the framework of probabilistic population codes, as long as the noise is small enough \cite{latham2003optimal}. Besides, the authors of \cite{burak2012fundamental}  also explored how noisy neural spiking drives the instantaneous attractor state to drift, which modeled how the stored memory in CANNs of probabilistically spiking neurons degrade over time by diffusion.  In accurate path integration tasks \cite{burak2009accurate}, it was shown that the CANN can `accurately integrate velocity inputs over a maximum of $\sim$10-100 meters and $\sim$1-10 minutes' under proper condition settings. Another interesting application is in studying the eye still effect \cite{seung1996brain}. It has been found that the eye slowly drifts to the center of the oculomotor range, so overcoming these perturbations and keeping its position is a complex task. The author modeled the problem using a point attractor and a line attractor (corresponding to the properties of two neuron populations respectively: one is to store the memory of the eye position and the other to read it out), and found the conditions for either attractors to be stable and robust against perturbations. 

In CANNs, different coupling profiles can give rise to different stabilization mechanisms. Neurons in multiple modules may have various kinds of couplings, which determine the dynamical behaviors of the whole network \cite{stringer2003self, stringer2002self, stringer2004self, stringer2005self, machens2008design, bush2014hybrid, kang2019geometric, amari1977dynamics}. In the early work by Amari, the network response is stabilized by rectangularly distributed couplings with a negative resting potential \cite{amari1977dynamics}. A more common approach uses Mexican hat couplings \cite{burak2009accurate}, namely, each neuron receives excitatory couplings from its surrounding neurons and inhibitory ones from those further away, so that the whole neural network is stabilized. Another commonly adopted coupling form is Gaussian coupling \cite{seeholzer2019stability, fung2012dynamical, wang2015rich, fung2015spontaneous}. Besides, there are also other nonlinear coupling forms such as exponential coupling \cite{burak2012fundamental}, cosine function coupling \cite{xie2002double}, and linear couplings shown in \cite{pettine2021excitatory}, which can lead to different responses by controlling their properties and strengths. Different coupling ranges directly determine the influence range. Different from the Mexican hat, Gaussian couplings from a neuron are always excitatory or inhibitory and therefore cannot stabilize the response bump alone. Here the network requires a global inhibition mechanism, in which neuron’s responses are stabilized via subtracting or dividing the summation of the responses of all neurons. As for the exponential couplings \cite{burak2012fundamental}, they enable the single neuron to have a long inhibitory range coupling with all other neurons, and the strength of coupling depends on the distance between neurons, which depends only on varying inhibitions within the neural network to stabilize the response dynamics. Incidentally, the continuous attractors are also learned widely in machine learning area \cite{xiang2021coexistence}. Besides, people also explore the performance of CANNs without recurrent excitation \cite{boucheny2005continuous}, whose results are consistent with the experimental observations in thalamic head direction cells. Despite the variation in details, the common stabilization mechanism of the bump profiles involves short-range excitation and long-range or global inhibition.

Based on the interaction between different modules, the brain is able to integrate its collected information to get a comprehensive picture of the surroundings \cite{fetsch2013bridging, stanford2005evaluating}. Multisensory information processing was investigated extensively in areas such as visual-auditory \cite{jaekl2007auditory}, visual-vestibular  \cite{zhang2016decentralized} or other kinds of combinations  \cite{ernst2002humans}. Thus, the purpose of this paper is to give a comprehensive picture of the dynamical and static properties of bimodular CANNs in a broad range of parameters such as reciprocal couplings, input strengths and disparities, which have not been done systematically before. In this paper, we focus on the study of the bimodular CANN structure, simulating the dynamics of the network and exploring their behaviors in multisensory information processing due to the interactions between two neural modules. Compared with single-module CANNs, bimodular networks are able to process information coming from different sensory modalities separately or simultaneously. As shown in this paper, by applying distinct inputs, the responses of the bimodular CANNs are very diverse. Furthermore, the couplings between the two neural modules play vital roles during information processing, either responding to two static stimuli or tracking a moving stimulus in one modality and a static stimulus in the other \cite{fung2009tracking, fung2015fluctuation, wang2015rich}.

The rest of the paper is organized as follows. After describing the network architecture in Section \ref{SectionII}, we discuss in Section III the network response when the inputs are static, and in Section IV when one input is static and the other moving. Section V consists of a comparison of network behaviors in the presence and absence of recurrent couplings. The paper is concluded in Section VI. The Appendix contains calculations of bump positions and heights providing supporting results in the main text.

\section{Network Architecture}
\label{SectionII}
\subsection{Single Layer CANNs}
\begin{figure}[htbp]
\centering
\includegraphics[width=0.46\textwidth]{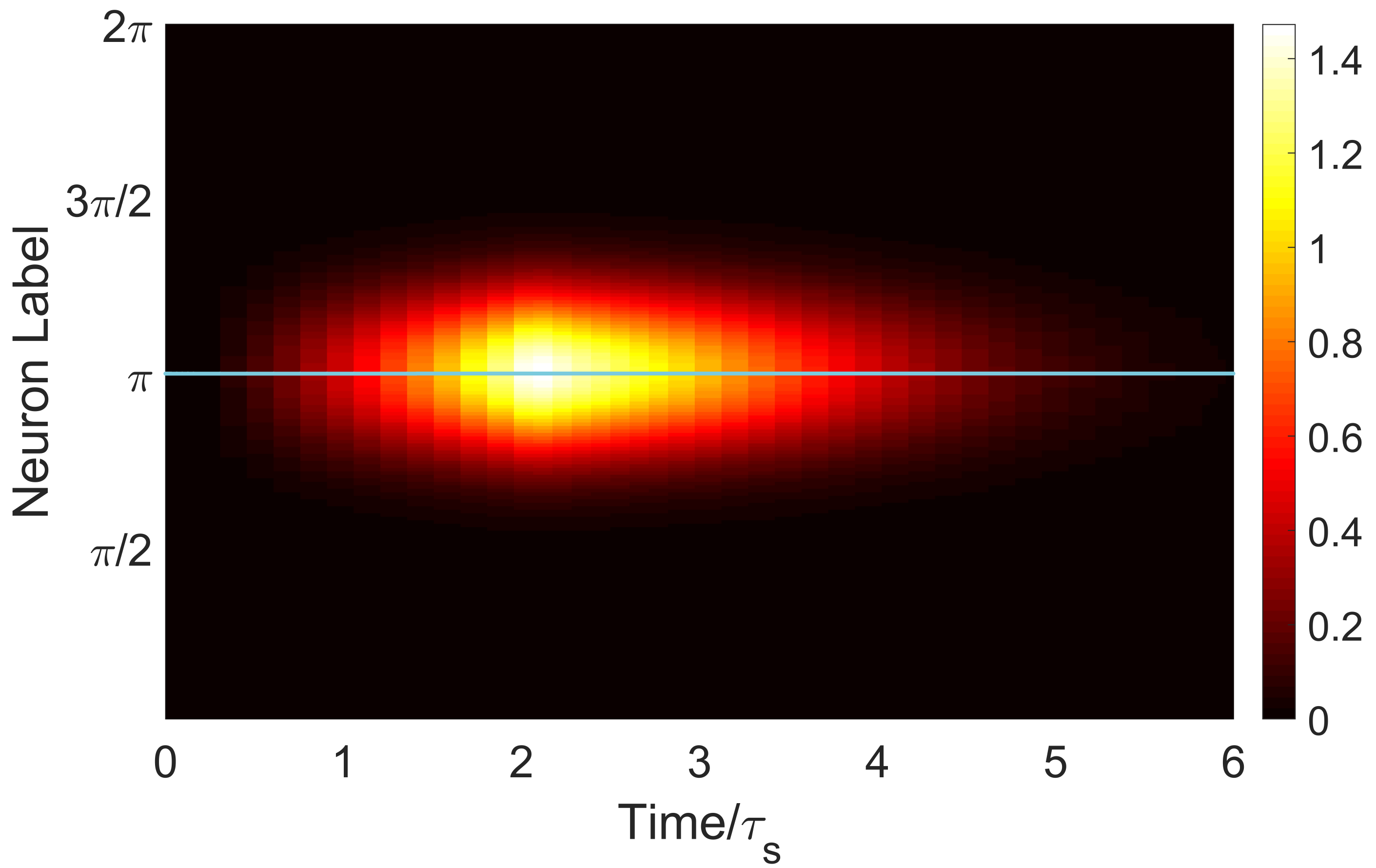}
\caption{The responses of a unimodular CANN when the external input is only applied during 0 to 2$\tau_s$. The responses gradually fade after the input is removed. Light blue lines indicate the input positions ($\pi$). }
\label{fgsilent}
\end{figure}

We first describe single-module CANNs which process a one-dimensional stimulus which can be regarded as the position or moving direction of an object, or head direction or other continuous variables. Each neuron in the network has its own preferred stimulus (direction), and the preferred stimuli of all neurons in the network cover the whole range of stimulus. 
Therefore, the population of neurons in the network can encode all possible values of the stimulus. Denoting $U(x,t)$ as the synaptic input received by the neuron whose preferred stimulus is $x$ at time $t$, and $x$ $\in$ [0, 2$\pi$). The dynamics of the $U(x,t)$ is \cite{fung2008dynamics,fung2010moving,fung2012dynamical}

\begin{equation}
\tau \frac{\partial U(x,t)}{\partial t} = -U(x,t) +\rho \int_{-\infty}^\infty J(x,x') r(x',t)dx' + I_{ext}(x,t) ,
\label{eq1}
\end{equation}

\noindent
where $\tau$ is the time constant of the  synaptic input, controlling the rate at which the synaptic input converges, typically at the order of 10 ms \cite{dayan2003theoretical}. $I_{ext}(x,t)$ denotes the external input to the neuron preferring stimulus $x$ at time $t$, and 
$\rho$ is the density of neurons covering the stimulus range.  The couplings between the neurons preferring $x$ and $x'$ are represented by Gaussian functions $J(x,x')$:

\begin{equation}
J(x,x') = \frac{J_0}{\sqrt{2\pi}a} \exp \left[ -\frac{(x-x')^2}{2a^2} \right],
\label{eq2}
\end{equation}

\noindent
where $a$ defines the interaction range among the neurons. It can be seen from Eq. (\ref{eq2}) that the coupling depends on the displacement $x-x'$ (modulo $2\pi$ for angular variables) and hence is 
translationally invariant. This endows the network with the ability to support  a continuous family of attractors. The function $r(x,t)$ denotes the firing rate at time $t$ and position $x$:
\begin{equation}
r(x,t) = \frac{[U(x,t)]_+^2}{1+k \rho \int_{-\infty}^\infty  [U(x',t)]_+^2 dx'},
\label{eq3}
\end{equation}

\noindent
in which $[U]_+ \equiv \max(U,0)$, and $k$ is the global inhibition, which controls the extent to which the firing rate saturates \cite{carandini2012normalization}. For  $0<k<k_c\equiv J_0^2 \rho/(8\sqrt{2 \pi}a)$ and for $a\ll 2\pi$, the CANNs can support a continuous family of stationary states \cite{fung2010moving}. 
To model neurons in the sensory cortex, in this study, $k=1.1k_c$ which means there is no sustained activity without external inputs, as shown in Fig. \ref{fgsilent} where the responses decay to zero after switching off external inputs.

To simplify, we rescale the parameters: $\widetilde{U}= \rho J_0 U$, $\widetilde{I}_{ext} = \rho J_0 I_{ext}$, $\widetilde{r} = (\rho J_0)^2 r$, $\widetilde{k} = \frac{8 \sqrt{2\pi}ak}{\rho J_0^2}$. Then Eq. (\ref{eq1}) and Eq. (\ref{eq3}) can be rewritten as:

\begin{equation}
\tau \frac{\partial \widetilde{U}(x,t)}{\partial t} = -\widetilde{U}(x,t) + \int_{-\infty}^\infty\frac{J(x,x')}{J_0} \widetilde{r}(x',t)dx' + \widetilde{I}_{ext}(x,t) ,
\label{eq4}
\end{equation}
\begin{equation}
\widetilde{r}(x,t) = \frac{[\widetilde{U}(x,t)]_+^2}{1+\frac{\widetilde{k}}{8\sqrt{2\pi}a} \int_{-\infty}^\infty dx' [\widetilde{U}(x',t)]_+^2}.
\label{eq5}
\end{equation}

\subsection{Bimodular CANNs}
\begin{figure}[htbp]
\centering
\includegraphics[width=0.5\textwidth,height = 0.350\textwidth]{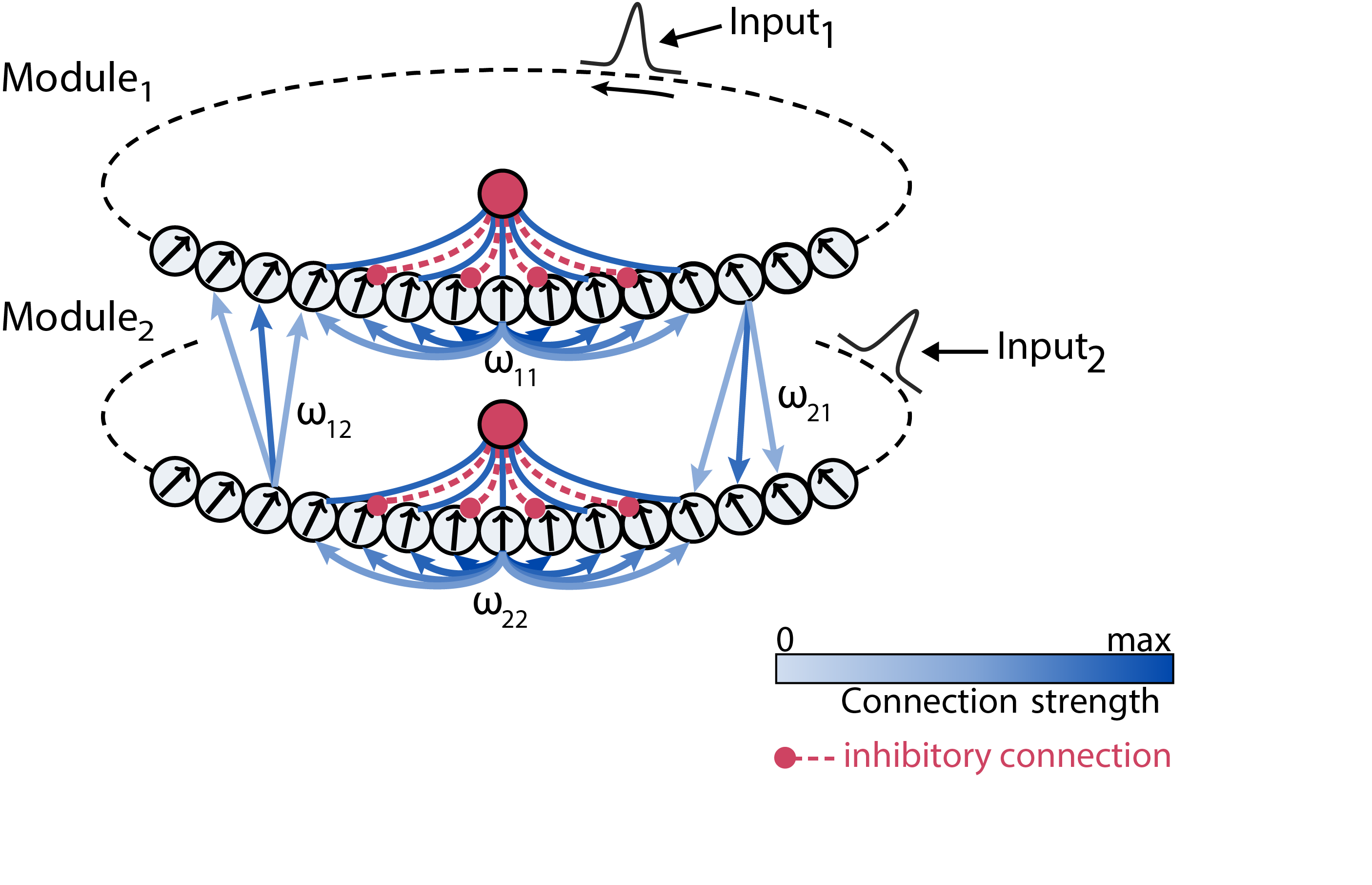}
\caption{The bimodular CANNs architecture. }
\label{fg1}
\end{figure}

A single CANN only models a hypercolumn of neurons in a brain area, whereas the cortex is composed of many hypercolumns which are coupled with each other.
These coupled hypercolumns can be located within the same brain area, or even from different areas which process different sensory modalities~\cite{felleman1991distributed,shams2008benefits,gu2008neural,dokka2015multisensory,molholm2004multisensory,fetsch2012neural,wallace2004revised}.
Therefore, we generalize the single-module CANN to a bimodular structure through adding couplings between the two modules in the bimodular CANNs, consistent with previous network models (e.g., \cite{zhang2012neural,zhang2016decentralized,piech2013network,li1998neural}). 
The network architecture is shown in Fig. \ref{fg1}. For  simplicity, we consider the case that the neurons are evenly distributed in the two modules.  Each neuron has its own preferred stimulus, indicated by the arrows in the neurons. Generalized from Eq. (\ref{eq4}), the dynamical equations of the bimodular CANNs model are (for convenience, we use $U$, $r$, $k$, $I$, $J(x,x')$ to denote the rescaled variables $\widetilde{U}$, $\widetilde{r}$, $\widetilde{k}$, $\widetilde{I}$, $J(x,x')/J_0$ in the rest of the paper):
 
\begin{align}
\tau \frac{\partial U_1(x,t)}{\partial t} &= -U_1(x,t) + \omega_{11}\int_{-\infty}^\infty J_{11}(x,x') r_1(x',t)dx' \nonumber \\
& + \omega_{12}\int_{-\infty}^\infty J_{12}(x,x') r_2(x',t)dx' + I_{1ext}(x,t) , \nonumber \\
\tau \frac{\partial U_2(x,t)}{\partial t} &= -U_2(x,t) + \omega_{22}\int_{-\infty}^\infty J_{22}(x,x') r_2(x',t)dx' \nonumber \\
& + \omega_{21}\int_{-\infty}^\infty J_{21}(x,x') r_1(x',t)dx' + I_{2ext}(x,t) . 
\label{eq6}
\end{align}

The recurrent coupling strength within module 1 (module 2) is denoted as $\omega_{11}$ ($\omega_{22}$), and their strengths are fixed at 1.0 in this paper. The coupling from module 1 (module 2) to module 2 (module 1) is denoted as $\omega_{21}$ ($\omega_{12}$). As for the coupling functions between two modules, we continue to adopt Gaussian functions similar to that in Eq. (\ref{eq2}):
\begin{equation}
J_{ij}(x,x') = \frac{1}{\sqrt{2\pi}a} \exp \left[ -\frac{(x-x')^2}{2a^2} \right], i,j \in \{1, 2, i \neq j \}.
\label{eqpJ}
\end{equation}
\noindent

The firing rates  (responses) in each module of bimodular CANNs are given by
\begin{align}
&r_1(x,t) = \frac{[U_1(x,t)]_+^2}{1+\frac{k}{8\sqrt{2\pi}a} \int_{-\infty}^\infty dx' [U_1(x',t)]_+^2}, \nonumber \\
&r_2(x,t) = \frac{[U_2(x,t)]_+^2}{1+\frac{k}{8\sqrt{2\pi}a} \int_{-\infty}^\infty dx' [U_2(x',t)]_+^2}.
\end{align}

There are also two external inputs $I_{1ext}$  and $I_{2ext}$ to the two modules respectively.  In this paper, both of the external stimuli are  in Gaussian forms:
\begin{align}
I_{1ext}& = I_{01} \exp \left[-\frac{(x-z_1)^2}{4a^2} \right], \nonumber   \\
I_{2ext}& = I_{02} \exp \left[-\frac{(x-z_2)^2}{4a^2} \right].
\end{align}
\noindent
$I_{01}$ and $I_{02}$ denote the magnitudes of external inputs respectively, and $x$ denote the positions of neurons.  The central positions of inputs are denoted as $z_1$ and $z_2$ respectively, which are either static or moving.

\begin{figure*}[htbp]
\centering
\includegraphics[width=1\textwidth]{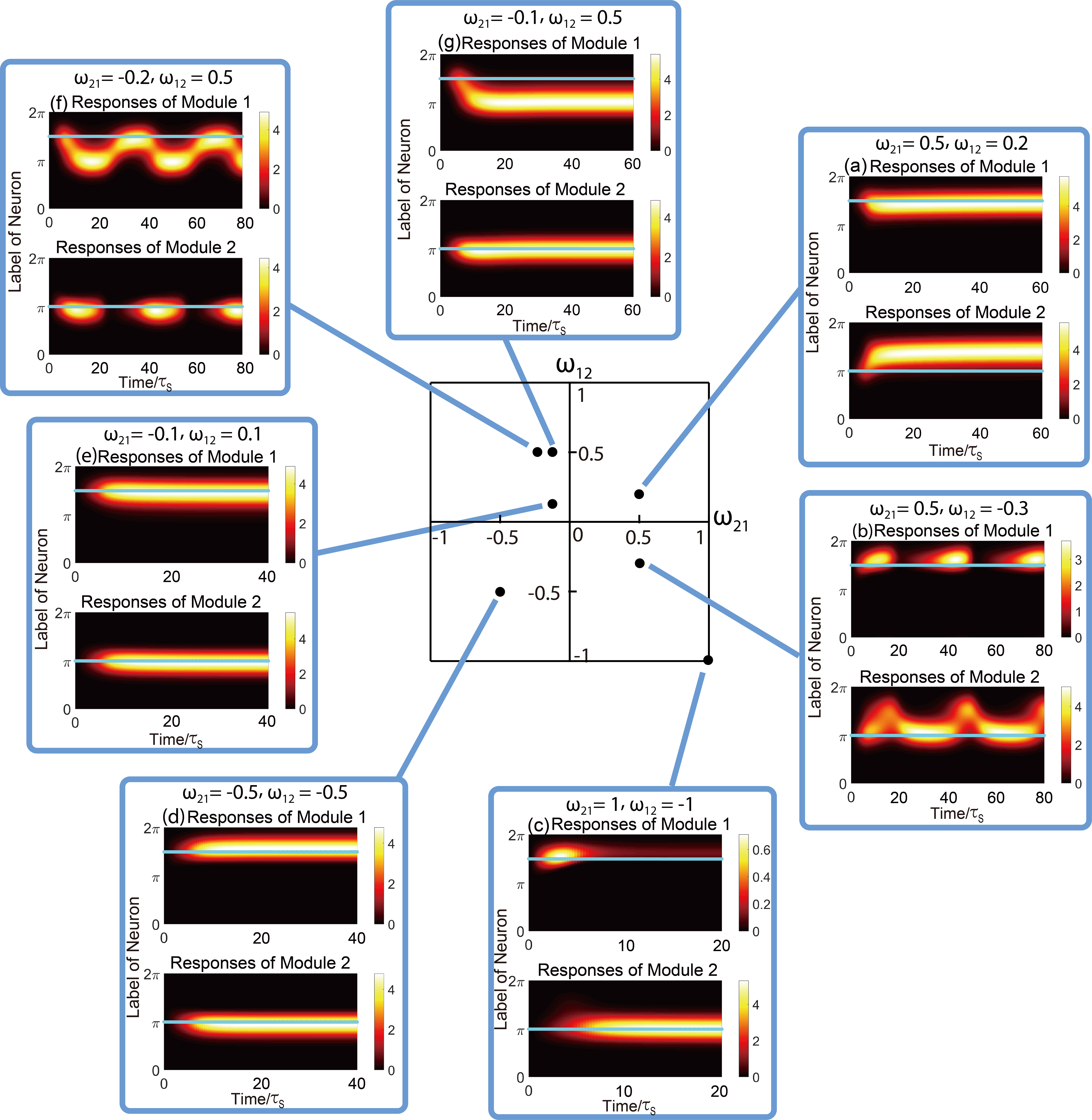}
\caption{The firing rates (responses) of bimodular CANNs when various inter-modular couplings are imposed. 
In each insert the top (bottom) plot shows the population responses of 1st (2nd) network module.
External inputs $I_{01} = I_{02} = 0.7$. Other parameters: $\omega_{11} = \omega_{22} = 1, a = 0.5, k = 1.1$. $I_{1ext}$ is fixed at position $3 \pi/2$, and $I_{2ext}$ is fixed at position $\pi$. Light blue  lines indicate trajectories of inputs. }
\label{fgsum}
\end{figure*}

\section{Static Inputs}
\label{Sec3}
\subsection{Dependence on Inter-Modular Couplings}
\label{SecA}
Different inter-modular couplings can give rise to various behaviors of the neural network as shown in Fig. \ref{fgsum}. 
We divide this spectrum of behaviors into three major types which are summarized in below. Mechanistically, in bimodular networks the bump in the module which receives excitatory inter-modular couplings is attracted towards the bump in the efferent module, whereas those receiving inhibitory couplings are repelled.

{\it (1) Coexistence of persistent bumps}: This can be found in regimes where the inter-modular couplings are both excitatory (Fig. \ref{fgsum}(a)) or both inhibitory (Fig. \ref{fgsum}(d)), or one being excitatory and another being weakly inhibitory (Figs. \ref{fgsum}(e) and \ref{fgsum}(g)). The positions of the bumps are shifted relative to the external input due to the inter-modular interactions, and the amount of shifts increases with the strength of the inter-modular couplings, as evident in Fig. \ref{fgsum}(a) where the bump in module 2 is shifted more  than that in module 1 due to $\omega_{21} > \omega_{12}$.

Similarly, in Fig. \ref{fgsum}(d), the bumps in both modules are shifted away from the external inputs due to the repulsion effect of the inter-modular couplings.

Figures \ref{fgsum}(e) and \ref{fgsum}(g) illustrate the case that one inter-modular coupling is excitatory while the reciprocal is weakly inhibitory. The bump in module 1 is attracted towards input 2 due to the excitatory $\omega_{12}$, and the bump in module 2 is pushed away from its own input due to the inhibitory $\omega_{21}$. The two cases differ in their behaviors of bump heights, which will be analyzed in further details in the next subsection.

{\it (2) Population spikes} (Figs. \ref{fgsum}(b) and \ref{fgsum}(f)): Population spikes refer to the periodic on-and-off population responses, which occur when one inter-modular coupling is excitatory and another is \emph{moderately} inhibitory. For example in Fig. \ref{fgsum}(b), after the bumps have been built up in both modules, the bump in module 1 is pushed away from its input position, meanwhile, the bump in module 2 is attracted by module 1. As module 2's bump moves toward the bump in module 1, it sends stronger inhibition to module 1 (due to reduced disparity of the two bumps) and eventually silence module 1's responses. Once the module 1 is silent, the attraction experienced by the bump in module 2 vanishes accordingly, causing it to return to the input position. In turn, the inhibition on the bump in module 1 weakens, allowing it to build up again. 

As these population spikes are temporal modulations of the population neural activity,  they are also known as ensemble synchronizations, representing extensively coordinated rises and falls in the discharge of many neurons \cite{loebel2002computation, holcman2006emergence}. In the past, the proposed mechanism of population spikes is the presence of short-term synaptic depression (STD), referring to the reduction of synaptic efficacy of a neuron after firing due to the depletion of neurotransmitters \cite{dayan2003theoretical, stevens1995facilitation, markram1996redistribution}. As far as we know, this is the first time an alternative mechanism is proposed for the formation of population spikes based on an excitatory-inhibitory feedback loop. In fact, it has been shown that STD and inhibitory feedback loops play many similar roles in the dynamical behaviors of neural networks such as anticipative tracking \cite{fung2015fluctuation}, and the generation of population spikes adds to the list of these similarities.

Localized population spikes, such as those illustrated in Figs. \ref{fgsum}(b) and \ref{fgsum}(g), have the ability to encode spatial and temporal information. It was a proposed mechanism to explain resolution enhancement as observed in transparent motion experiments in the middle temporal (MT) area of the nervous system \cite{treue2000seeing}. In these experiments, the neural system was found to be able to resolve motion disparities narrower than the tuning width of the neurons. The proposed mechanism based on population spikes is that two sequences of population spikes are generated in response to two external inputs with narrow angular separation \cite{fung2013resolution}.

{\it (3) Momentary spikes} (Fig. \ref{fgsum}(c)): This occurs when one inter-modular coupling is excitatory and the reciprocal one is \emph{strongly} inhibitory. As shown in Fig. \ref{fgsum}(c), the responses in module 1 are inhibited severely soon after the initiation of the input. It has a minor effect of attracting the bump in module 2 towards input 1, but the bump in module 2 quickly stabilizes and remains effectively unaffected by module 1, due to the responses in module 1 being inhibited too quickly. The momentary behavior in module 1 is useful in coding transient information of sensory inputs.

\subsection{Excitatory and Displacement Effects on Bump Heights}
\label{SecB}

Since the bump heights can encode the statistical weight of the carried information \cite{deneve1999reading}, it is important to study how they are determined in the coupled network module.
The bump heights are dependent on two effects. The first one is the excitatory effect. In general, the bump in the module receiving excitatory inputs from another module is higher than its counterpart that receives inhibitory inputs. This is shown in Fig. \ref{fgsum}(g), in which the bump in module 1 is higher than that in module 2.

\begin{figure}[htbp]
\centering
\includegraphics[width=0.45\textwidth]{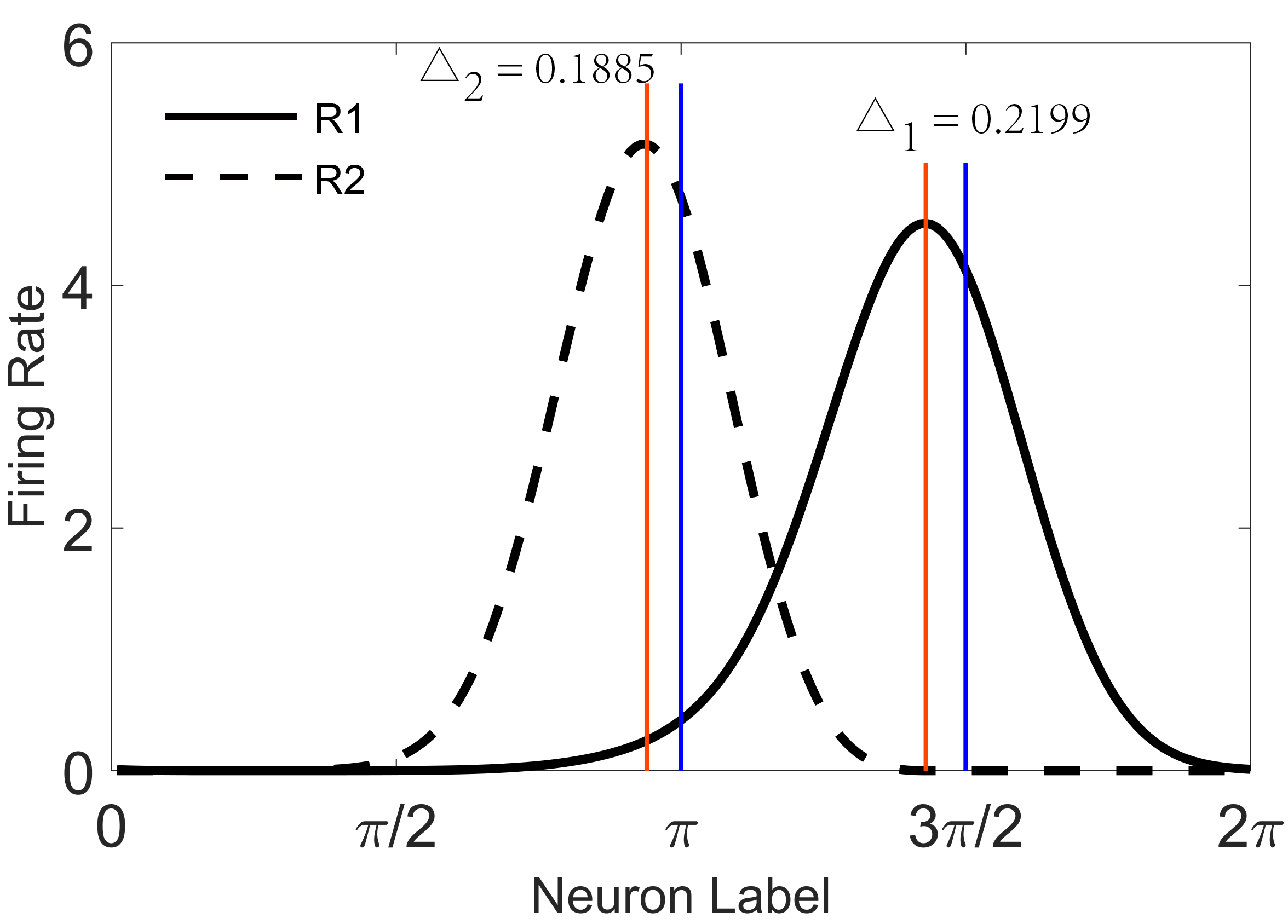}
\caption{The displacement effect on bump height. The bumps in both modules are displaced by inter-modular interaction. The inter-modular couplings: $\omega_{12} = 0.1$, $\omega_{21} = -0.1$. Other parameters: $\omega_{11} = \omega_{22} = 1, a = 0.5, k = 1.1$. $R_1$ denotes responses of module 1, and $R_2$ denotes responses of module 2. The magnitudes of $I_{01}$ and $I_{02}$ are 0.5. Inputs 1 and 2 are applied at $1.5\pi$ and $\pi$, respectively. The displacement between input position (blue lines) and peak position of response (red lines) of each module is denoted and marked by $\Delta_1 (\Delta_2)$. }
\label{fgs1}
\end{figure}

On the other hand, the second effect becomes operative and may produce an opposite effect on the bump height. This second effect is the displacement effect due to the inter-modular coupling, which is more significant for weak inputs. Due to the inter-modular couplings, the peak positions of the bumps in both modules are either attracted or repelled from the respective stimulus positions. When the disparity is within the range of the inter-modular couplings, the tendency to displace the bumps increases with the disparity \cite{fung2010moving}. This displacement weakens the efficacy of the input stimuli and results in a reduction of the amplitude. As shown in Fig. \ref{fgsum}(e), the bump in module 1 receiving excitatory inter-modular coupling has weaker responses compared with that in module 2 receiving inhibitory inter-modular coupling. The displacement effect in Fig. \ref{fgsum}(e) is magnified in Fig. \ref{fgs1} when the input strengths are further reduced to 0.5. This is contrary to the prediction of the excitatory effect and the displacement of the bump position away from its external input needs to be considered. 

The remaining issue is why the bump in module 1 receiving excitatory inter-modular coupling is displaced more than that in module 2 receiving inhibitory inter-modular coupling. To see this, we consider the displacement of the bump center $x_1$ in module 1 with respect to the input $z_1$. As derived in Appendix in the limit of weak inter-modular couplings,

\begin{equation}
x_1 - z_1 = \frac{H_{12}(I_{02}-H_{21})(z_2-z_1)}{I_{01}I_{02}-H_{12}H_{21}},
\label{neweq12}
\end{equation}
\noindent
where $H_{ij}=\omega_{ij}R_{0j}/\sqrt{2}$ is the inter-modular contribution from module $j$ to module $i$ at the  peak position (corresponding to the third term in the right-hand side of Eq. (\ref{eq6})), and $R_{0i}$ is the maximum firing rate of the bump in module $i$. Not surprisingly, the displacement is proportional to disparity $z_2 - z_1$. Furthermore, the displacement is proportional to the first term $H_{12}$. In the expression $I_{02} - H_{21}$, the first term shows that the bump displacement increases with the external input in module 2. The second term is a higher order effect in the limit of weak inter-modular coupling. It originates from the competition between input 2 and the inter-modular input from module 1 to 2. When the inter-modular input is excitatory ($\omega_{21} > 0$), the competition weakens the effective strength of input 2, since the inter-modular input tends to divert the bump in module 2 from input 2 to input 1 when  the external inputs have disparity. On the other hand, when the inter-modular input is inhibitory, the effective strength of input 2 increases.

Hence, while the bumps in both modules are displaced by inter-modular interaction, the one in module 1 has a larger displacement since it receives an inter-modular input from module 2 reinforced by $\omega_{21}$ being inhibitory. In contrast, the bump in module 2 has a smaller displacement since it receives an inter-modular input from module 1 weakened by $\omega_{12}$ being excitatory. This explains the lower bump height in module 1. However, as also shown in Fig. \ref{fgs1}, the displacement effect is of a higher order of the inter-modular coupling strength and is much weaker than the excitatory effect.

\subsection{Bimodular Networks as Decoders of Multisensory Integration}
\label{SecC}

Understanding the behaviors of the bimodular network is highly relevant to the process of information integration in modularized neural systems \cite{zhang2016decentralized}. It is known that human neural systems are able to integrate sensory inputs from multiple channels in Bayes-optimal ways \cite{ernst2002humans,knill2004bayesian,pouget2013probabilistic}. In probabilistic population coding, the mean and reliability (e.g., the inverse of the variance for Gaussian distributions) of the posterior distribution of the stimuli after receiving the cues are inferred by the center-of-mass position and the height of the bumps, respectively \cite{ma2006bayesian,beck2008probabilistic, pouget1998statistically, wu2001population,vasudeva2016inference}. For bimodular networks, the bump in each module represents the marginal posterior distribution of its corresponding stimulus. In the literature the Bayes-optimal performance of the bimodular networks has been illustrated by Gaussian distributions of the priors and likelihoods \cite{zhang2016decentralized, zhang2019complementary,ernst2002humans, sato2007bayesian}. 

Here, we extend the Bayesian analysis to include both excitatory and inhibitory reciprocal couplings of bimodular networks, and study how they encode multisensory information. For both cases, we consider the prior given by
\begin{equation}
p(s_1, s_2) \propto \exp \left[- \frac{(s_1 - s_2)^2}{2\sigma_s^2} \right].
\label{neweq13}
\end{equation}

For correlated priors, $\sigma_s^2$ is positive. For anti-correlated priors, $\sigma_s^2$ is negative and will lead to divergence of the total prior probability. Hence, we will restrict our consideration to the case that $s_1$ and $s_2$ are bounded. Nevertheless, as we shall see, when the likelihood distributions are sharp enough, the posterior probabilities remain well defined in the region well within the bounds of $s_1$ and $s_2$. With the likelihoods given by
\begin{equation}
p(z_i|s_i) \propto \exp \left[- \frac{(z_i - s_i)^2}{2\sigma_i^2} \right], i = 1, 2, 
\label{neweq14}
\end{equation}
\noindent
we arrive at the marginal posterior distribution
\begin{equation}
p(s_1|z_1, z_2) = 
p_{\max}(s_1|z_1, z_2) 
\exp \left[ -\frac{(s_1 -  \langle s_1|z_1, z_2 \rangle)^2}{2V(s_1|z_1, z_2)} \right],
\label{neweq15}
\end{equation}
\noindent
where 
\begin{equation}
\langle s_1|z_1, z_2 \rangle = \frac{\sigma_1^{-2}z_1 + \sigma_{2s}^{-2}z_2}{\sigma_1^{-2} + \sigma_{2s}^{-2}}; \quad 
\sigma_{2s}^2 \equiv \sigma_2^2 + \sigma_s^2,
\label{neweq16}
\end{equation}
\begin{equation}
V(s_1|z_1, z_2) = \left[ \sigma_1^{-2} + \sigma_{2s}^{-2} \right]^{-1},
\label{neweq17}
\end{equation}
\begin{equation}
p_{\max}(s_1|z_1, z_2) = \frac{1}{\sqrt{2 \pi V(s_1|z_1, z_2)}}.
\label{neweq18}
\end{equation}
\noindent
A similar equation for the posterior mean $\langle s_2 | z_1,z_2 \rangle$ can be obtained by permuting the network labels 1 and 2. To relate the inference of a module to its direct input and the inference of the other module, we have
\begin{equation}
\langle s_1|z_1, z_2 \rangle = \frac{\sigma_1^{-2}}{\sigma_1^{-2} + \sigma_s^{-2}}z_1 + \frac{ \sigma_s^{-2}}{\sigma_1^{-2} + \sigma_s^{-2}}\langle s_2|z_1, z_2 \rangle. 
\label{neweq19}
\end{equation}
\noindent
On the other hand, the position of the bump in module 1 as derived in Appendix is given by
\begin{align}
x_1 &= \frac{I_{01} e^{\frac{-(x_1 - z_1)^2}{8a^2}}}{I_{01} e^{\frac{-(x_1 - z_1)^2}{8a^2}} + H_{12}e^{\frac{-(x_1 - x_2)^2}{8a^2}}}z_1  \nonumber \\
&+ \frac{H_{12} e^{\frac{-(x_1 - x_2)^2}{8a^2}}}{I_{01} e^{\frac{-(x_1 - z_1)^2}{8a^2}} + H_{12}e^{\frac{-(x_1 - x_2)^2}{8a^2}}}x_2.
\label{neweq20}
\end{align}

As will be presented below, the simplified version of this result is more compatible with the Gaussian nature of the priors and likelihoods in the Bayesian framework. Nevertheless, as described in Appendix, this result becomes compatible with likelihoods in which the variance of the Gaussian likelihood is a slowly broadening function. We will not pursue this modification further in this paper.

Provided that the disparity between the two cues is not excessively large compared with the synaptic range $a$ of the networks, this result can be approximated by the linear relation
\begin{equation}
x_1 = \frac{I_{01}}{I_{01} + H_{12}}z_1 + \frac{H_{12}}{I_{01} + H_{12}}x_2.
\label{neweq21}
\end{equation}

Comparing Eqs. (\ref{neweq21}) and (\ref{neweq19}), we note that the center-of-mass positions of the bumps in the two modules can be used to decode the posterior means of the stimuli in the Bayesian framework with Gaussian priors and likelihoods.

Furthermore, if the network has direct access to the input $z_1$, say, through a direct independent circuit, then in principle the network has sufficient information to decode the weight of the reliability $\sigma_1^{-2}$ of input 1 relative to the reliability $\sigma_s^{-2}$ of the prior from the ratio of the shifts $x_2-x_1$ to $x_1-z_1$.

We note that if $\sigma_s^2$ is positive, then the corresponding network will have positive values of $\omega_{12}$ and $\omega_{21}$.  This corresponds to the case that $\omega_{12}$ and $\omega_{21}$ encode correlated (attractive) priors. If $\sigma_s^2$ is negative, we only consider the case that the likelihood distributions are sharp enough such that $\sigma_1^2$, $\sigma_2^2 < -\sigma_s^2$, and thus the posterior distributions have well defined maxima. In this case, when $\sigma_1^2 + \sigma_2^2 > -\sigma_s^2$, there is no real solution for the variance in Eq. (\ref{neweq17}). When the likelihood distributions are further sharpened to satisfy $\sigma_1^2 + \sigma_2^2 \leq -\sigma_s^2$, the corresponding network will have negative values of $\omega_{12}$ and $\omega_{21}$, encoding anti-correlated (repulsive) priors.

The case of $\omega_{12}$ and $\omega_{21}$ taking up opposite signs can also be interpreted in the Bayesian framework with correlated likelihoods. To see this, we consider a generic prior described by Eq. (\ref{neweq15}) but the likelihood is given by
\begin{equation}
p(z_1, z_2|s_1, s_2) \propto \exp \left[- \frac{1}{2} \sum_{i, j =1}^2 (z_i - s_i) (C^{-1})_{ij} (z_j - s_j) \right],
\label{neweq22}
\end{equation}
\noindent
where $C_{ij} = \langle (z_i - s_i)(z_j-s_j) \rangle$. Using Bayes' rule, the marginal posterior probability is given by a Gaussian distribution with mean and variance given by
\begin{equation}
\langle s_1|z_1, z_2 \rangle = \frac{(C_{22}-C_{12}+\sigma_s^2)z_1 + (C_{11}-C_{12})z_2}{C_{11}+C_{22}-2C_{12}+\sigma_s^2},
\label{neweq23}
\end{equation}
\begin{equation}
V(s_1|z_1, z_2) = \frac{C_{11}C_{22} - C_{12}^2 + C_{11}\sigma_s^2}{C_{11}+C_{22}-2C_{12}+\sigma_s^2}.
\label{neweq24}
\end{equation}
\noindent
In the context of bimodular networks, we have
\begin{equation}
\langle s_1|z_1, z_2 \rangle = \frac{\sigma_s^2}{\sigma_s^2+C_{11}-C_{12}}z_1 + \frac{C_{11}-C_{12}}{\sigma_s^2+C_{11}-C_{12}} \langle s_2|z_1, z_2 \rangle.
\label{neweq25}
\end{equation}
\begin{equation}
\langle s_2|z_1, z_2 \rangle = \frac{\sigma_s^2}{\sigma_s^2+C_{22}-C_{12}}z_2 + \frac{C_{22}-C_{12}}{\sigma_s^2+C_{22}-C_{12}} \langle s_1|z_1, z_2 \rangle.
\label{neweq26}
\end{equation}
\noindent
There exist likelihood functions in which $C_{11}-C_{12}$ and $C_{22}-C_{12}$ have opposite signs. For example, consider the likelihood being the covariance matrix
\begin{equation}
\left(
\begin{array}{cc}
C_{11} & C_{12} \\
C_{21} & C_{22} 
\end{array}
\right)
=
\left(
\begin{array}{cc}
3/4 & 1 \\
1 & 3/2 
\end{array}
\right).
\label{neweq27}
\end{equation}

This corresponds to the case of a rather strong correlation between the two likelihoods, with stimulus 2 being dominated by stimulus 1 (i.e., $C_{22} > C_{11}$). In this case, response 2 is attracted by response 1 due to the higher accuracy of the latter.

On the other hand, response 1 is repelled by response, due to its exceedingly strong reliance on input 1, and the repulsion by response 2 is merely the compensation (noting that the sum of the two pre-factors in Eq. (\ref{neweq23}) should be equal to 1). With $\sigma_s^2=1/2$, this results in
\begin{align*}
\langle s_1|z_1, z_2 \rangle & = 2z_1 - \langle s_2|z_1, z_2 \rangle, \\
\langle s_2|z_1, z_2 \rangle & = \frac{1}{2}z_2 + \frac{1}{2} \langle s_1|z_1, z_2 \rangle.
\end{align*}

The corresponding network structure would have negative and positive values for $\omega_{12}$ and $\omega_{21}$, respectively.


\subsection{Decoding Behavior of Bimodular Networks}
\label{SecD}

\begin{figure*}[htbp]
\centering
\includegraphics[width=0.9\textwidth]{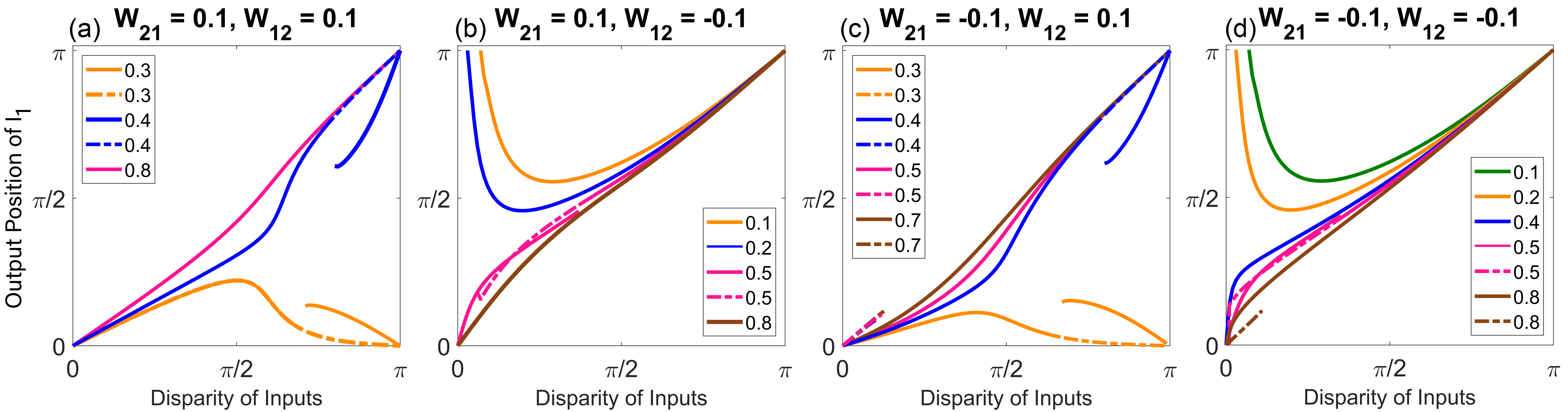}
\caption{The center-of-mass position of the responses in module 1 versus the input position of module 1 at different weak inter-modular couplings. Different colors indicate different $I_{01}$ of Input 1. Input 2 is  fixed at position 0 and $I_{02} = 0.7$. $\omega_{11} = \omega_{22} = 1, a = 0.5, k = 1.1$.  Solid lines: steady states reached from the all-quiet initial condition. Dashed lines: steady states reached from other initial conditions. Values of  $I_{01}$ are indicated in the legend. (a) Inter-modular couplings $\omega_{21} = \omega_{12} = 0.1$. (b) Inter-modular couplings $\omega_{21} = 0.1, \omega_{12} = -0.1$. (c) Inter-modular couplings $\omega_{21} = -0.1, \omega_{12} = 0.1$. (d) Inter-modular couplings $\omega_{21} = -0.1, \omega_{12} = -0.1$.}
\label{newbutterfly}
\end{figure*}

We further study how the center-of-mass of the bump depends on the disparity when the network stays at the all-quiet state initially (Fig. \ref{newbutterfly}). 
We vary the position and the amplitude of $I_{1ext}$, but fix the amplitude and position of $I_{2ext}$ to a sufficiently strong value to make the network exhibit competition effects (thus the horizontal axis is the disparity of the two inputs).

(1)	{\it Model averaging versus model selection}: It is worthwhile to summarize our observations from the point of view of model averaging versus model selection. These concepts are frequently encountered in the studies of causal inference \cite{shams2010causal}, but they are also relevant to the present work on multisensory integration, which is often considered as the precursor to causal inference in downstream neural modules \cite{cuppini2017biologically}. Causal inference refers to the process of inferring whether or not an event A is caused by another event B. When the neural system receives two different signal cues, it needs to determine whether they come from a common source. If they are from one source, the two cues will be integrated during processing, otherwise they will be dealt with separately. So, inferring whether two input cues have a common source is a crucial step for information processing. `Model selection' and `model averaging' are two possible strategies in inferring the underlying causal structures (e.g., in hierarchical causal inference models \cite{shams2010causal, wozny2010probability}). In causal inference, there are two hypotheses proposed: a common cause hypothesis and separate causes hypothesis, and each is weighted by their probabilities. `Model averaging' means that the estimate of the stimuli is the weighted average of the estimates derived from the two different hypothesized causal structures. As for `model selection', it means that the network chooses the most probable one as the  final decision, instead of performing weighted averaging \cite{shams2010causal, wozny2010probability}.
In normative models of causal inference for two channels using a model averaging strategy, cues from these channels are integrated at low disparity, resulting in an averaged prediction as described in the previous section. However, when the disparity is high, the stimuli of the individual channels are inferred to be independent, and the channel with the highest posterior probability will be inferred, that is, a model selection strategy is adopted.

(2) {\it Module 1 fed by excitatory inter-modular couplings} ($\omega_{12} > 0$) (Figs. \ref{newbutterfly}(a) and \ref{newbutterfly}(c)): The `butterflies' in Fig. \ref{newbutterfly} show similar behaviors as those in causal inference. It is worth noting that the network behaviors at low and high disparity are dramatically different. When the disparity is not large (around $<0.5 \pi$ in Fig. \ref{newbutterfly}(a) and $<0.4 \pi$ in Fig. \ref{newbutterfly}(c)), the bump position is a linearly interpolated function of $z_1$ and $z_2$ as predicted by Eq. (\ref{neweq16}). The slope is effectively constant, indicating that a model averaging strategy is adopted. The bump position in module 1 is closer to input 1 as the amplitude of input 1 increases.

In contrast, when the disparity is large, the output position of module 1 either drifts towards $z_1$ when $I_{1ext}$ is strong. or towards $z_2$ when $I_{1ext}$ is weak. This is the regime in which the network operates in the model selection mode.

(3) {\it Module 1 fed by inhibitory inter-modular couplings} ($\omega_{12} < 0$) (Figs. \ref{newbutterfly}(b) and \ref{newbutterfly}(d)): In this case, the behavior depends on the strength of input 1. When the input strength is weak ($I_{01} \leq 0.3$), the output is repelled to distant positions due to the inhibitory couplings. This corresponds to the case that the denominators on the right-hand side of Eq. (\ref{neweq21}) is negative, and models the situation in Bayesian analysis that the likelihoods are not sufficient to overcome the repulsive prior. Due to the inhibitory effects from module 2, the firing rate of these diverged bumps are very low.

When the strength of input 1 is sufficiently strong, the bump in module 1 becomes strong enough to resist the repulsion from module 2 and maintains its position around input 1. We also find a linear regime at low disparity (output position $ <0.2 \pi$). The slope of the linear regime is greater than 1, and decreases with the strength of input 1 as the repulsive effect from module 2 is less significant. This agrees with the Bayesian analysis in Eq. (\ref{neweq16}) for repulsive priors, resembling the model averaging strategy. At higher disparity, output 1 effectively tracks input 1, consistent with the model selection strategy.

(4) {\it Multiple steady states}: We also observe the existence of multiple steady states, and the eventual steady state depends on the initial condition. When the network is initialized to an all-quiet condition, the bump positions typically reach a steady position which is in between $z_1$ and $z_2$ when $\omega_{21} > 0$. The firing rate profile of these steady states show that two states coexist in the network, with one near $I_{1ext}$ or $I_{2ext}$  and the other staying between (see $I_{01} = 0.4$ and $0.3$ in the high disparity regime of Figs. \ref{newbutterfly}(a) and \ref{newbutterfly}(c)). Other steady states exist at high disparity, but they need to be reached from initial conditions other than the all-quiet state.

When $\omega_{21} < 0$, we note that two steady states exist at intermediate strengths of $I_{1ext}$ (see $I_{01} = 0.5$ in Figs. \ref{newbutterfly}(b) and \ref{newbutterfly}(d)). At low disparity, the  final steady state accessed from the all-quiet initial condition has a low firing rate. When the disparity increases, the repulsion effect of input 2 weakens and the firing rate undergoes a discontinuous transition. The bump position in module 1 also undergoes a discontinuous transition but the discontinuity is not large.

When module 1 repels module 2 ($\omega_{21} < 0$), an extra steady state also exists at low disparity and very strong $I_{01}$ (see $I_{01} = 0.7$ in Fig. \ref{newbutterfly}(c) and $I_{01} = 0.8$ in Fig. \ref{newbutterfly}(d)). In this state, the response in module 1 is dominated by its own input, and the output position is effectively the same as cue 1. At the input strengths indicated in Figs. \ref{newbutterfly}(c) and \ref{newbutterfly}(d), these states cannot be accessed from the all-quiet initial condition, but at higher values of  $I_{01}$, they are so dominant that they can be reached from the all-quiet initial state.

In summary, the bimodular networks are able to decode the information described by the Gaussian prior and likelihood in the Bayesian framework. However, it should be cautioned that the Bayesian decoding performance may be hindered by the existence of multiple steady states, limiting the range of validity in the space of input disparity and input strengths.

(5) {\it Bias}: An alternative view of the decoding process is to consider the bias, which quantifies the extent of influence of input 2 on output 1. In the present context, it is the change of output position 2 from the corresponding real input position divided by the disparity of the two inputs, namely, $(x_1-z_1)/(z_2-z_1)$. The bias is positive when input 1 is attractive, and negative when repulsive, and is expected to be independent of the disparity in the model averaging regime. We will focus on states accessible from all-quiet initial conditions at low disparity.

\begin{figure*}[htbp]
\centering
\includegraphics[width=\textwidth]{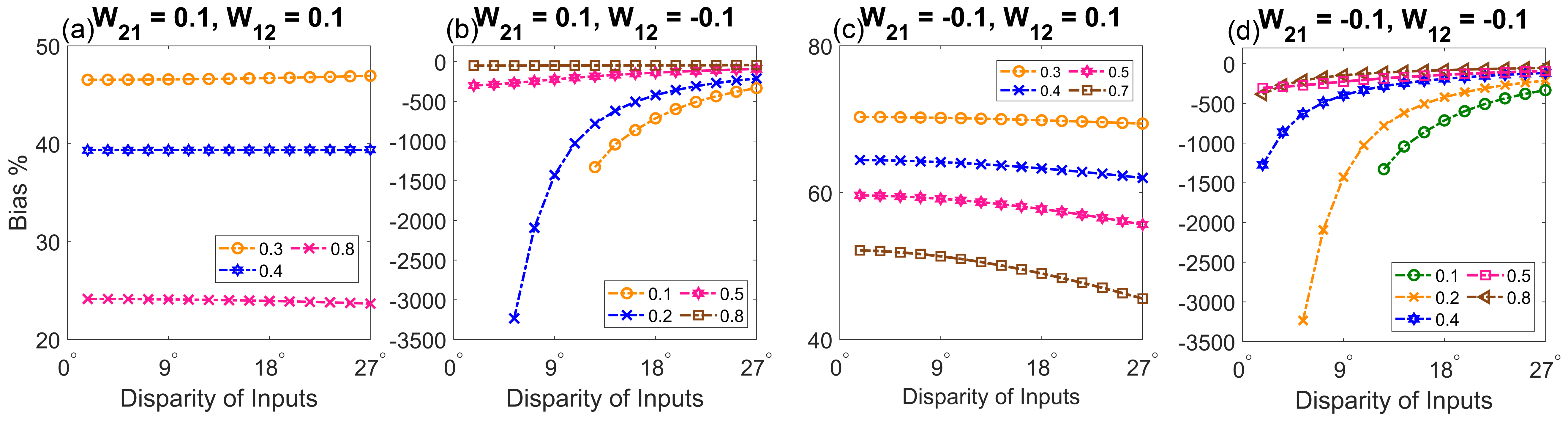}
\caption{Module 1 Perception Bias as a function of disparity. Different colors indicate different $I_{01}$ of Input 1. Input 2 is  fixed at position 0 and $I_{02} = 0.7$. $\omega_{11} = \omega_{22} = 1, a = 0.5, k = 1.1$.  Solid lines: steady states reached from the all-quiet initial condition. Dashed lines: steady states reached from other initial conditions. Values of  $I_{01}$ are indicated in the legend. (a) Inter-modular couplings $\omega_{21} = \omega_{12} = 0.1$. (b) Inter-modular couplings $\omega_{21} = 0.1, \omega_{12} = -0.1$. (c) Inter-modular couplings $\omega_{21} = -0.1, \omega_{12} = 0.1$. (d) Inter-modular couplings $\omega_{21} = -0.1, \omega_{12} = -0.1$.}
\label{bias}
\end{figure*}

As shown in Figs. \ref{bias}(a) and \ref{bias}(c) in which module 1 is fed by excitatory inter-modular couplings, the bias is positive, indicating that the decoded position of input 1 is attracted by input 2. The values of the biases are effectively constant in the range of low disparity, as expected in the model averaging regime. The biases decay at higher disparity, which is not shown in the panels. Comparing Figs. \ref{bias}(a) and \ref{bias}(c), we observe that the bias in the latter is higher for the corresponding inputs. This is due to the position of output 2 being repelled further away from module 1, thus allowing output 1 to be attracted further towards input 2.

As for Figs. \ref{bias}(b) and \ref{bias}(d) in which module 1 is fed by inhibitory inter-modular couplings, the bias is negative and is effectively constant in the range of low disparity, provided that input 1 is not too weak. When input 1 is weak, the bias diverges to large negative values. These results agree with the predictions of Bayesian analysis. We observe that the biases in Fig. \ref{bias}(d) is not as negative as those in Fig. \ref{bias}(b) for the corresponding inputs, due to mutual inter-modular repulsion.

The effectively constant and positive biases in Fig. \ref{bias}(a) are consistent with those reported in previous literature. For example, Cuppini {\it et al}. considered a network structure with a visual module and an auditory module reciprocally connected, and their outputs are projected to a common causal inference area, and biases with a gradual decrease with disparity are reported \cite{cuppini2017biologically}. On the other hand, when the statistics of the biases are separately evaluated when the causal inference system identifies a common cause or different causes, the biases were reported to be positive for a common cause, and negative for different causes \cite{kording2007causal}. In contrast, we have not included a downstream circuitry for causal inference in our study, and will leave this issue for future studies.

\subsection{Time Delay Effects}
\label{SecE}

\begin{figure}[htbp]
\centering
\includegraphics[width=0.4\textwidth]{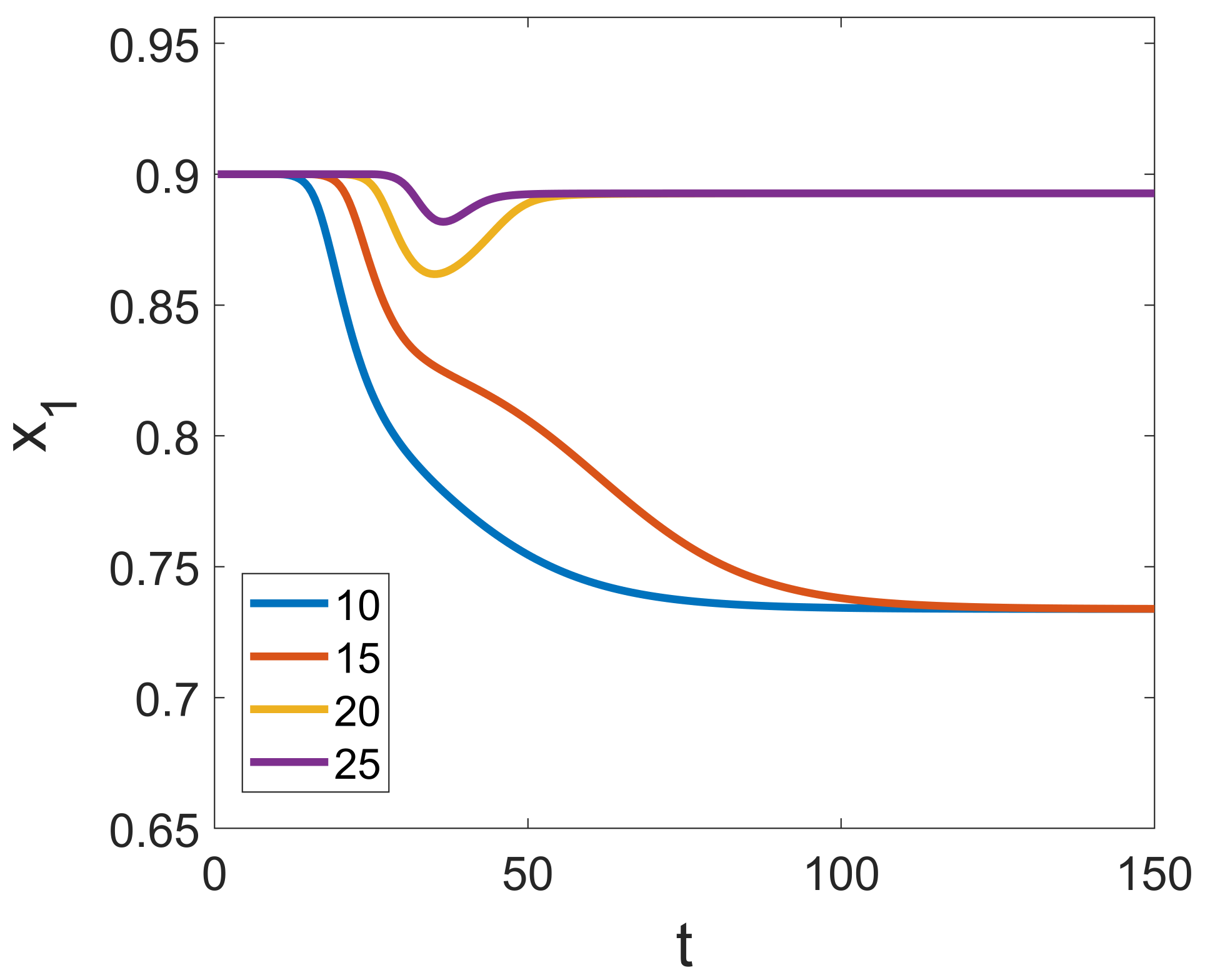}
\caption{ The time evolution of the center-of-mass position of module 1 in a bimodular network with $\omega_{21} = \omega_{12} = 0.1$, $I_{01}=0.4$, $I_{02}=0.7$,  $\omega_{11} = \omega_{22} = 1, a = 0.5, k = 1.1$, $z_1 = 0.9 \pi$ and $z_2 = 0$, lying in the regime with two steady states in Fig. \ref{newbutterfly}(a), and $x_1 = 0.734 \pi$ and $0.893 \pi$. Input 1 is switched on at $t = 0$, and input 2 is switched on at $t = 10, 15, 20$ and $25$ ms, as indicated by different colors.}
\label{timedelay}
\end{figure}

Due to the existence of multiple steady states in the high disparity regime, the final state reached by the network is dependent on the initial condition. Particularly relevant to modular neural networks is the processing times in different modules. In Fig. \ref{timedelay} we model situations in the regime where the network has two steady states but a stimulus onset asynchrony (SOA) is present. As indicated in the high-disparity regime in Fig. \ref{newbutterfly}(a), the steady state with the low value of $x_1$ in module 1 is more stable under the inter-modular influence from module 2, and hence module 1 converges to the low value of $x_1$ when both inputs are switched on at the same time. However, when the onset of input 2 is delayed, the bump in module 1 has sufficient time to build up around input 1, and the state with higher $x_1$ is established. Figure \ref{timedelay} illustrates how the competition between the two attractors evolve with increasing time delay. These time delay effects are relevant to the processing of visual and auditory signals, in which visual signals are delayed by an order of $10^2$ ms [25], so that the inference in the auditory channel is not required to be excessively strong for it to remain independent of the visual channel in the regime of high disparity.

\section{Moving Inputs}

\subsection{Dynamical Behaviors}
\begin{figure*}[htbp]
\centering
\includegraphics[width=0.8\textwidth,height=0.443\textwidth]{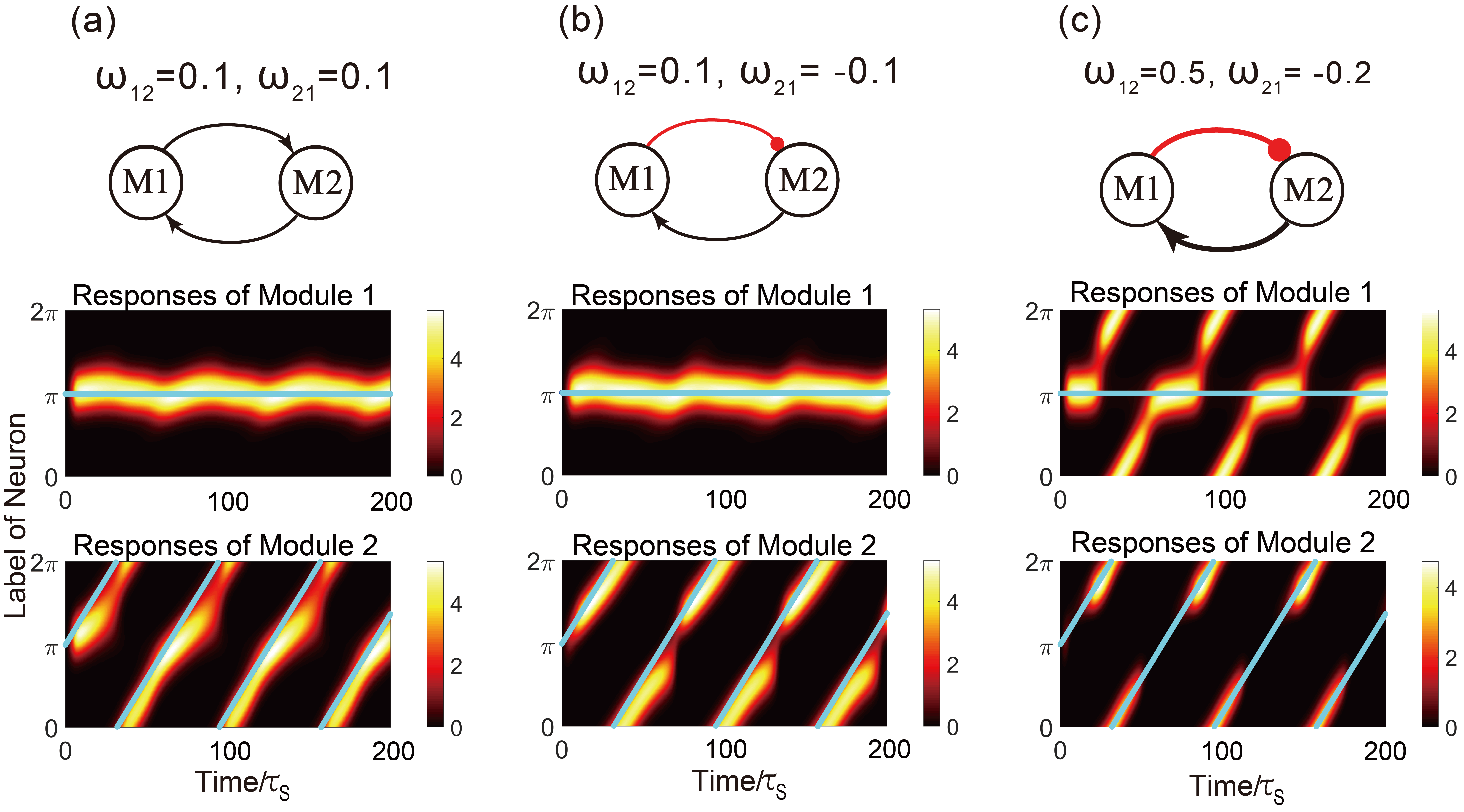}
\caption{Behavior of bimodular CANNs under different weak inter-modular couplings and stimuli, one static and one moving. $I_{1ext}$ is static and located at position $\pi$. $I_{2ext}$ is a moving stimulus, moving velocity $v_2 = 0.01$ rad/ms. Light blue lines indicate the inputs' trajectories. $k = 1.1$ and $I_{01} = I_{02} = 0.7$ except $I_{01} = 1.4$ in (c). In each of (a), (b) and (c), the top panel shows the sketches of the network structure with $M1$ and $M2$ denoting modules 1 and 2 respectively. The middle panel shows the firing rates (responses) of module 1, and the bottom panel shows the firing rates (responses) of module 2.}
\label{fg4}
\end{figure*}

The brain receives temporally dynamical inputs, and hence it is imperative to study the processing of moving stimuli in bimodular CANNs. In unimodular networks, a rich spectrum of behaviors is already observed \cite{ben1997traveling}. Following the previous section, we consider external inputs of moderate strengths. 
As an example, we consider the case that the  network receives a static input 1 but a moving input 2 with constant speed (Fig. \ref{fg4}). Both inputs are applied to the network simultaneously from time $t\ge 0$ (input positions are indicated by light blue lines, Fig. \ref{fg4}). 

As Figs. \ref{fg4}(a) and \ref{fg4}(b) show, the responses in module 1 are oscillating around the input 1 position owing to the excitatory inter-modular couplings $\omega_{12}$ from module 2 to module 1. In Fig. \ref{fg4}(a), the responses of module 2 at position around $\pi$ are enhanced, resulting from the positive couplings $\omega_{21}$, whereas in Fig. \ref{fg4}(b), the $\omega_{21}$ is inhibitory, thus the responses of module 2 are inhibited around the position $\pi$.

Figures \ref{fg4}(a)-\ref{fg4}(b) show that the static and moving stimuli can interact with each other via the couplings between two modules. However, the competition between the external inputs 1 and 2 is not obvious in module 1 since input 1 is direct and input 2 works via inter-modular couplings. In order to explore the competition, we present a relatively extreme situation, shown in Fig. \ref{fg4}(c).

Figure \ref{fg4}(c) shows the dynamics of the bimodular CANNs under a stronger excitatory inter-modular coupling ($\omega_{12}$) accompanied by a weaker inhibitory coupling ($\omega_{21}$). The input magnitude of the static input is much stronger than that of the moving input. Now,  the dynamics of the network is totally different from that in Fig. \ref{fg4}(b). In Fig. \ref{fg4}(c), on account of the strong inhibitory couplings from module 1, the responses in module 2 around the position $\pi$ are almost fully suppressed. Only after input 2 has moved away or before it arrives at the position $\pi$ can stable and strong responses be built in module 2. However, influenced by the strong attraction from module 2, the responses in module 1 can only sustain its static state when the responses in module 2 are inhibited. When the responses in module 2 are rebuilt, they again attract the responses in module 1, inducing module 1 to track the moving $I_{2ext}$ instead of its own static input $I_{1ext}$.

\subsection{Phase Diagrams}

To obtain a more comprehensive picture, we introduce the tracking mean square deviations with respect to the static and moving inputs as references. A comparison of their magnitudes reveals whether the responses are tracking the static or moving inputs. Below, we denote the modules receiving static and moving inputs as modules $s$ and $m$ respectively. Since module $m$ receives a moving stimulus, we particularly focus on the mean square deviations in module $m$,

\begin{align}
\sigma_s^2 &= \langle (x_m(t) - v_s t)^2 \rangle_t - \langle x_m(t) - v_s t \rangle_t^2, \nonumber \\
\sigma_m^2 &= \langle (x_m(t) - v_m t)^2 \rangle_t - \langle x_m(t) - v_m t \rangle_t^2,
\label{eq29}
\end{align}
\noindent
where $x_m(t)$ denotes the center of mass of the responses in module $m$, and $v_m$ and $v_s$ indicate the moving velocities of two external inputs with $v_s = 0$. $\langle \cdots \rangle_t$ represents average over time. $\sigma_m^2$ and $\sigma_s^2$ denote the mean square deviations of the responses in module $m$ with respect to the two external input positions of the network, respectively. When $\sigma_s^2$ is less than $\sigma_m^2$, it means the module $m$ is tracking the static input more than its own moving input. Otherwise, it tracks the moving input more.

\begin{figure*}[htbp]
\centering
\includegraphics[width=\textwidth]{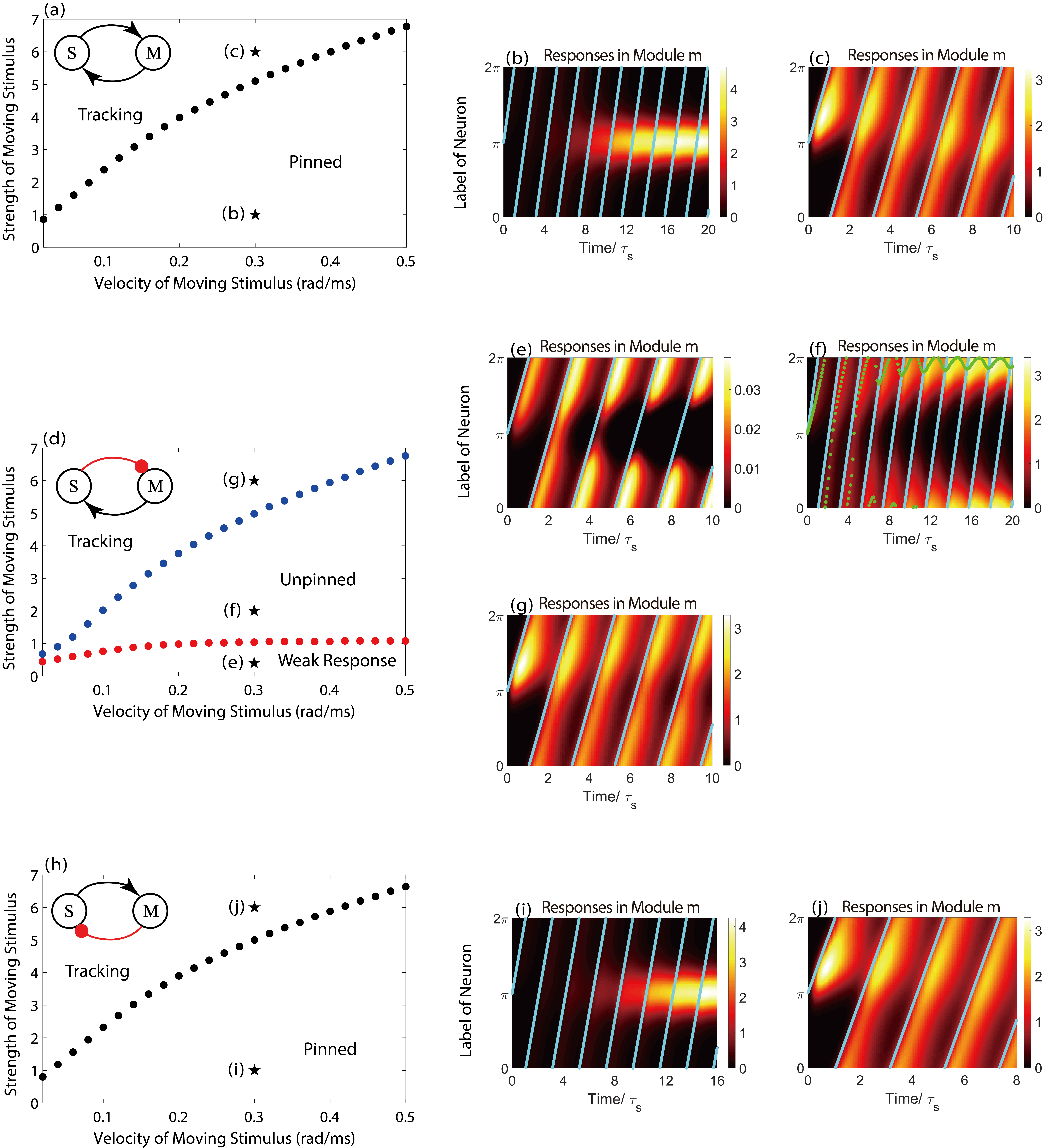}
\caption{The phase diagrams of the dynamical behaviors in module $m$ with moving stimulus ((a), (d) and (h)) and network behaviors at selected locations. The static input $I_{1ext}$ is fixed at the amplitude of 0.7, applied at position $\pi$, and the amplitudes of all couplings are fixed at 0.1. The trajectories of input 2 is indicated by light blue lines.  (b), (c):  The firing rates (responses) of the bimodular CANN in (a) when $I_{2ext}$ moves at a speed of 0.3 rad/ms, and the magnitude of $I_{2ext}$ is 1 and 3 respectively. (e), (f) and (g): The firing rates (responses) of the bimodular CANN in (d) when $I_{2ext}$ moves at a speed of 0.3 rad/ms, and the magnitude of $I_{2ext}$ is 0.4, 2 and 6 respectively. (i) and (j): The firing rates (responses) of the bimodular CANN in (h) when $I_{2ext}$ moves at a speed of 0.3 rad/ms, and the magnitude of $I_{2ext}$ is 1 and 6 respectively.}
\label{fg6}
\end{figure*}

Figure \ref{fg6} shows the phase diagrams of tracking behaviors. Three kinds of couplings are listed at the upper left corners, respectively. We also pick some points with the same moving velocities, but various moving input strengths as examples of the responses in module $m$ shown in Fig. \ref{fg6}. The bump in module $s$ is effectively pinned to the static input in this parameter range and will not be shown. This does not contradict the moving bump in module 1 in Fig. \ref{fg4}(c) where the intermodular coupling is much stronger. The corresponding data points are marked in Figs. \ref{fg6}(a), \ref{fg6}(d) and \ref{fg6}(h) respectively by black stars.

As shown in Figs. \ref{fg6}(b) and \ref{fg6}(i), the module $m$ cannot track its own moving stimulus when the stimulus is relatively weak, or the input moves too fast. The response is pinned by the static input. This is referred to as the pinned phase. As the moving velocity increases, stronger moving stimulus is needed to overcome the static interactions from the other module. When the moving input strength is sufficiently strong, module $m$ is able to catch up with the moving input. This is the tracking phase with $\sigma_m^2  < \sigma_s^2$ (see Figs. \ref{fg6}(c) and \ref{fg6}(j)). In Figs. \ref{fg6}(a) and  \ref{fg6}(h), in which module $s$ excites module $m$, the phase boundaries are similar, with the pinned phase at low strength of the moving input (see Figs. \ref{fg6}(b) and \ref{fg6}(i)) and the tracking phase at high strength (see Figs. \ref{fg6}(c) and \ref{fg6}(j)).

On the other hand, when module $s$ inhibits module $m$, the phase boundaries in Fig. \ref{fg6}(d) are different from the other two cases and an unpinned phase exists at intermediate strength of the moving input. In Fig. \ref{fg6}(d), module $m$ cannot build up stable and strong responses when the moving stimulus is very weak. This is referred to as the weak response phase. Furthermore, due to the inhibitory inter-modular couplings $\omega_{ms}$ and the weak moving input, the responses are suppressed temporarily when the moving bump passes by the inhibitory static input (see Fig. \ref{fg6}(e)). This region of temporary suppression even extends slightly beyond the boundary of the weak response phase.

As the strength of the moving stimulus increases, module $m$ is able to build strong and stable responses. Due to the repulsion by the static input, the bump is repelled from the static input and drift with a low velocity, resulting in the unpinned phase. The drift velocity has the same direction as that of the moving input, with the bump attracted forward towards the moving input when the latter is ahead, or attracted backward towards the moving input when the latter is behind (see Fig. \ref{fg6}(f), where the green curve indicates the drifting motion of the center of mass). The bump motion is heavily affected by the presence of the static input, which forms a barrier to the bump motion, causing the drift of the bump to slow down and reverse from forward attraction to backward attraction.

As the moving input strength continues to increase, the bump trajectory follows the moving input closer. When the input is strong enough, the bump is able to catch up with the moving input, and the network enters the tracking phase (see Fig. \ref{fg6}(g)) with $\sigma_m^2 < \sigma_s^2$. 

\section{Comparison with models without recurrent interactions}

An important ingredient in the ability of bimodular CANNs to track external inputs is the presence of recurrent couplings which have been studied extensively in both experiments and neural models \cite{voges2012complex, okun2015diverse, peron2020recurrent, sweeney2020population}. 
In comparison, multisensory integration has been traditionally analyzed using models with no recurrent connections between excitatory neurons~\cite{ma2006bayesian,ohshiro2011normalization}. 
Although the family of models without recurrent couplings can simulate some neural responses well and simple to analyze, the recurrent couplings and nonlinearities in the CANNs give rise to a rich spectrum  of population activities \cite{wang2015rich} and generate more accurate predictions \cite{gucclu2017modeling}.

\begin{figure*}[htbp]
\centering
\includegraphics[width=0.8\textwidth]{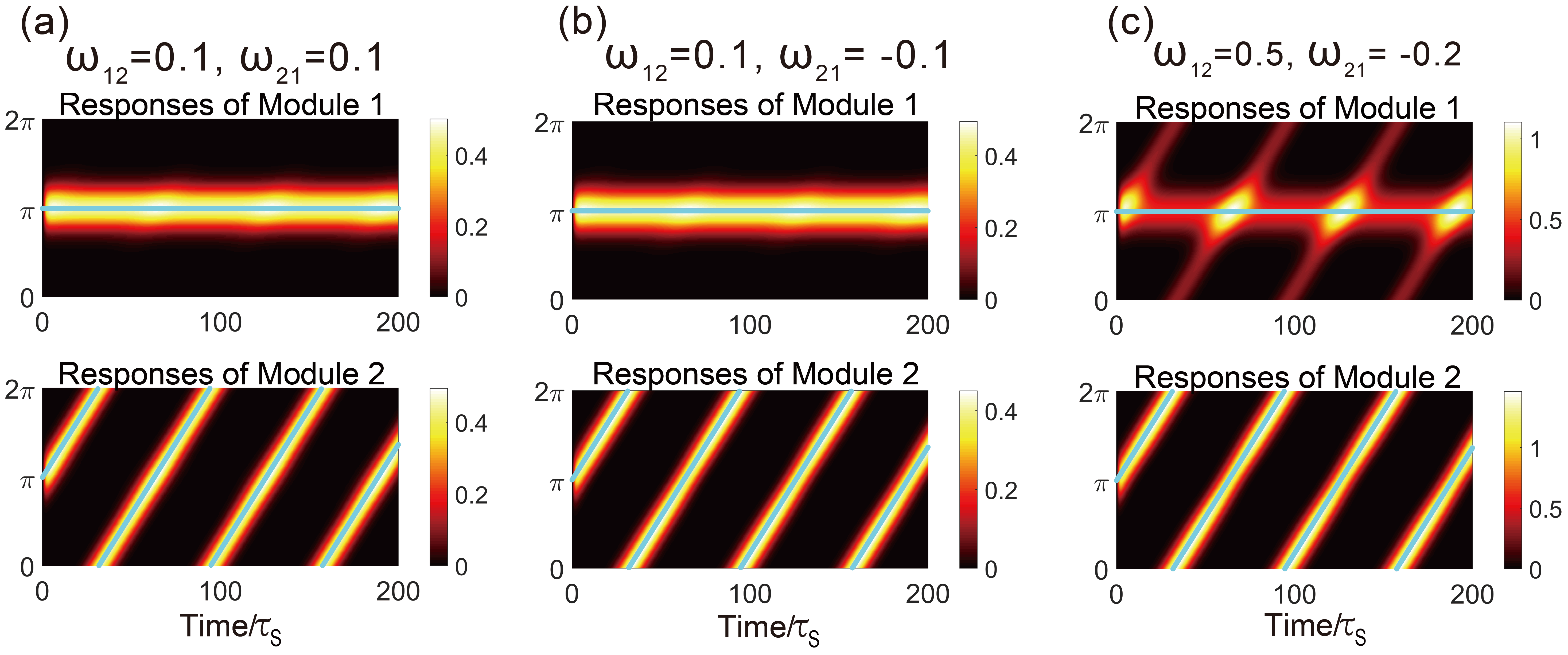}
\caption{The firing rates (responses) of the traditional model under different weak inter-modular couplings and stimuli, one static and one moving, which are indicated by light blue lines. (a) The same parameter settings as Fig. \ref{fg4} (a). (b) The same parameter settings as Fig. \ref{fg4} (d). (c) The same parameter settings as Fig. \ref{fg4} (g).}
\label{classicfg8}
\end{figure*}

We compare the dynamics of bimodular CANNs with models without recurrent couplings within each network module.
To make a fair comparison, we only remove the recurrent couplings within the CANNs and keep other parameters unchanged. 
We will use Figs. \ref{fg4} and \ref{classicfg8} as examples and apply the same stimuli to the classical neural network. The following four effects can be observed.

{\it 1) Bump height and displacement:} Removing the recurrent couplings weakens the influences on bump height and displacement from the other module (compare Fig. \ref{fg4} with Fig. \ref{classicfg8}). For example, as shown in Figs. \ref{classicfg8}(a) and \ref{classicfg8}(b) in which module 1 receives excitatory couplings ($\omega_{12}$) from module 2, only slight inter-modular influences in the height and displacement of the bump can be seen from the responses in module 1 when both stimuli are located at the central position $\pi$, where the responses in module 1 are barely enhanced. 

On the other hand, as shown in Figs. \ref{classicfg8}(a) and \ref{classicfg8}(b) module 2 receives either excitatory or inhibitory couplings from module 1 ($\omega_{21}$), but due to the lack of recurrent couplings, the inter-modular influences cannot spread out to other neurons, let alone the whole neural network. Therefore, the height and displacement of the bump in module 2 are almost not influenced by the inter-modular couplings. 

As for Fig. \ref{classicfg8}(c), since the inter-modular couplings are strong, the effects are most evident when the two stimuli are located at $\pi$, while the impacts are still slight compared with the dynamics in Fig. \ref{fg4}(c).

{\it 2) Unimodal/multimodal responses}: 
Removing recurrent connections within each network module causes the firing rate profile to split into two bumps (Fig. \ref{classicfg8}(c)), which completely changes the firing rate profile. 
In contrast, the bump remains single and intact in Fig. \ref{fg4}(c) due to the self-sustaining interactions of the recurrent couplings. 

\begin{figure*}[htbp]
\centering
\includegraphics[width=0.8\textwidth]{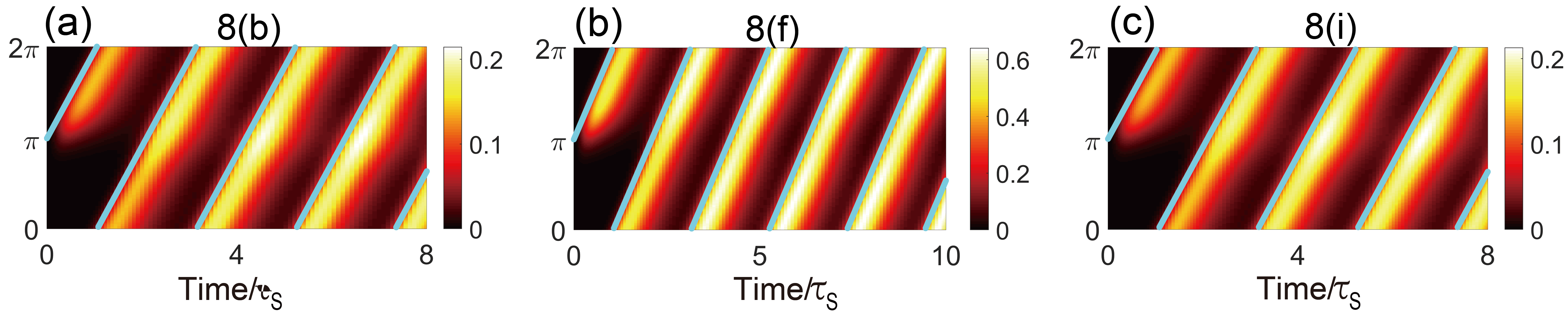}
\caption{The firing rates (responses) of the traditional model under different weak inter-modular couplings and stimuli, one static and one moving, indicated by light blue lines. The figures with the same parameter settings are indicated in the title of each panel.}
\label{classicfg9}
\end{figure*}

{\it 3) Response time}:  Comparing Figs. \ref{classicfg9}(a)-(c) with their corresponding figures in Fig. \ref{fg6}, we can see that the bumps are built up faster. This is because their bumps are lower and thus it takes a shorter time to build.

{\it 4) Mobility}: Comparing Figs. \ref{classicfg9}(a)-(c) with their corresponding figures in Fig. \ref{fg6}, we also see that the bumps have a higher  tendency to move.

In summary, we can see that network modules with and without recurrent couplings have different responses to external stimuli. Recurrent couplings have the effect of reducing the mobility of the responses and maintaining the unimodal nature of the direct input during its integration process with other indirect inputs, but the process of bump formation may require a longer duration. It is plausible that recurrent couplings are more important in later stages of information processing whereas they play a lesser role in the earlier information pathway.

\section{Conclusion}

We have generalized the study of unimodular CANNs to bimodular CANNs, endowing the network with the capacity to incorporate two sensory modalities. The inter-modular couplings in bimodular CANNs play important roles in determining the statics and dynamics of the network. Excitatory inter-modular couplings result in enhancing and attracting the responses of the efferent module, while inhibitory inter-modular couplings lead to suppressing and repelling effects for both static and moving inputs. The network behavior is determined by the interplay of the input strengths, their disparity, speed (for moving inputs) and the inter-modular couplings. 

 The most interesting case is the bimodular CANN with a reciprocal pair of excitatory and inhibitory inter-modular couplings. For static inputs, the inhibitory inter-modular feedback may produce population spikes which have been shown to enhance resolution of inputs with narrow separation \cite{fung2013resolution}. It also exhibits the anomalous behavior with the inhibited module having stronger-than-expected output than the excited module, due to the uneven displacement of the bumps in the respective modules. For static and moving inputs to the excited and inhibited modules respectively, a series of drifting responses with continuous and discontinuous evolution occur when the moving input strength increases and finally arriving at the tracking phase.

We have shown that bimodular networks can serve as decoders of posterior probabilities in the Bayesian framework for multisensory integration. They provide a neural substrate pre-conditioned for causal inference, as they yield integrated outputs at low disparity and segregated outputs at high disparity. Bimodular networks with a pair of excitatory inter-modular couplings encode attractive priors, whereas those with an inhibitory pair encode repulsive priors. Bimodular networks with excitatory inter-modular couplings and inhibitory reciprocal ones can also be used to model causal inference in which one channel is subordinate to the other (that is, both channels are subject to correlated noises but the dominating one has a weaker noise amplitude). When the disparity of the two inputs lie within the synaptic range, the bump position is a linear combination of the direct input position and the decoded indirect input position, fully consistent with the Bayesian prediction with Gaussian priors and likelihoods.  However, when the disparity is high, the bump positions become independent, modeling the adoption of a model selection strategy. This behavior justifies the modular structure as a pre-conditioning circuit for causal inference. 

On the other hand, we found that multiple steady states exist when the disparity of the inputs lie outside the synaptic range, with one state nearer the direct stimulus and the other nearer the indirect stimulus. Starting from an initial quiet state, the final state attained by the network depends on the relative strengths of the inputs. The existence of multiple states is relevant to how the decoded network state depends on the arrival time delays of the direct and indirect inputs, and is expected to be observable in experiments.

There are other applications of bimodular networks that have not been discussed in this paper. Bimodular networks with inhibitory couplings are important components in models of competitive decision making \cite{wang2002probabilistic, wang2008decision, wang2013top}. Besides, using bimodular networks with dynamical inputs, we can model multisensory psychophysics experiments such as the motion-bounce illusion experiment in which visual perception of moving inputs are influenced by auditory inputs \cite{shams2000you, shams2002visual, shimojo2001sensory, watkins2006sound}.

Multisensory interactions have been an important issue which has been studied extensively. Figuring out how the brain processes multisensory signals is an important topic not only in modeling the functions of the brain, but also in technological applications of neural computation. It has been commonly recognized that excitatory couplings between modules are important when the brain deals with different channels of signals that are correlated \cite{shams2008benefits}, and the inhibitory couplings are important when the brain processes signals that are uncorrelated or anti-correlated \cite{zhang2016decentralized, zhang2016congruent}. There have been experiments finding `congruent' and `opposite' cells \cite{gu2008neural}, whose responses to signals with different disparities can be rather diverse in experiments integrating visual and vestibular signals in the monkey's brain. In a recently proposed model explaining the functions of the congruent and opposite cells in Bayes-optimal inference \cite{zhang2019complementary}, the inter-modular couplings play an important role. Recent work also showed that the network structure to achieve Bayes-optimal performance incorporating both excitatory and inhibitory couplings depends on the prior distribution of the signals \cite{wang2017prior}. While most of the studies focus on the steady-state behaviors of the neural system, our work shows that dynamical and temporal behaviors are also important, and the transient behaviors of the neural system may also be useful in conveying information between the sensory modalities. Experiments based on temporal integration, such as the moving-bounce illusion experiment, can also be designed to further study the multisensory information processing.

\subsection*{Acknowledgments} This work is supported by grants from the Research Grants Council of Hong Kong (grant numbers 16322616, 16306817, 16302419 and 16302619).

\clearpage
\onecolumngrid
\section*{Appendix: Calculation of the Bump Position at Low Disparity and Weak Inter-modular Coupling}

\numberwithin{equation}{section}
\setcounter{equation}{0}
\renewcommand{\theequation}{A\arabic{equation}}
\renewcommand{\thefigure}{S\arabic{figure}}
\setcounter{figure}{0}

The steady-state equation of the bump in module 1 is given by
\begin{equation}
U_1(x) = \omega_{11}\int_{-\infty}^{\infty} \frac{dx'}{\sqrt{2 \pi} a^2} e^{\displaystyle -\frac{(x-x')^2}{4a^2}} \frac{U_1(x')^2}{B_1} + \omega_{12}\int_{-\infty}^{\infty} \frac{dx'}{\sqrt{2 \pi} a^2} e^{\displaystyle -\frac{(x-x')^2}{4a^2}} \frac{U_2(x')^2}{B_2} + I_{01}e^{\displaystyle -\frac{(x-z_1)^2}{4a^2}},
\label{eqA1}
\end{equation}
\noindent
where $B_i = 1 + k \int_{-\infty}^{\infty} dx U_i(x)^2 / 8\sqrt{2 \pi a^2}$ for $i=1,2$. The equation for module 2 can be obtained by permuting the module indices. The approximate solution to the steady-state equation proposed in \cite{fung2010moving} is based on the perturbation expansion with the lowest order term of the form
\begin{equation}
U_i(x) = U_{0i}e^{\displaystyle -\frac{(x-x_i)^2}{4a^2}} {\rm for} \ i = 1, 2.
\label{eqA2}
\end{equation}
\noindent
$U_{0i}$ is the maximum synaptic input of the bumps. Evaluating the integrals in Eq. (\ref{eqA1}), we get
\begin{equation}
U_{01}e ^{\displaystyle -\frac{(x-x_1)^2}{4a^2}} \approx \frac{\omega_{11}U_{01}^2}{\sqrt{2}B_1} e ^{\displaystyle -\frac{(x-x_1)^2}{4a^2}} + \frac{\omega_{12}U_{02}^2}{\sqrt{2}B_2} e ^{\displaystyle -\frac{(x-x_2)^2}{4a^2}} +  I_{01}e^{\displaystyle -\frac{(x-z_1)^2}{4a^2}}. 
\label{eqA3}
\end{equation}
\noindent
The solution of $U_{0i}$ is obtained by projecting the equation to the height mode (that is, multiplying both sides of the equation by $e^{-(x-x_1)^2/4a^2}$ and integrating $x$),
\begin{equation}
U_{01} = \frac{\omega_{11}R_{01}}{\sqrt{2}}+\frac{\omega_{12}R_{02}}{\sqrt{2}} e ^{\displaystyle -\frac{(x_1-x_2)^2}{8a^2}} + I_{01}e^{\displaystyle -\frac{(x_1-z_1)^2}{8a^2}},
\label{eqA4}
\end{equation}
\noindent
where $R_{0i}$ is the maximum firing rate of the bump in module $i$, and is given by
\begin{equation}
R_{0i} = \frac{U_{0i}^2}{B_i}, \quad B_i = 1 + \frac{kU_{0i}^2}{8}.
\label{eqA5}
\end{equation}
\noindent
The solution of $x_i$ is obtained by projecting the equation to the height mode (that is, multiplying both sides of the equation by $(x-x_1) e^{-(x-x_1)^2/4a^2}$ and integrating $x$),
\begin{equation}
0 = \frac{\omega_{12}R_{02}}{\sqrt{2}}(x_1-x_2) e ^{\displaystyle -\frac{(x_1-x_2)^2}{8a^2}} + I_{01}e^{\displaystyle -\frac{(x_1-z_1)^2}{8a^2}}.
\label{eqA6}
\end{equation}
\noindent
This results in the bump position given by 
\begin{equation}
x_1 = \frac{ I_{01}e^{\displaystyle -\frac{(x_1-z_1)^2}{8a^2}}}{ I_{01}e^{\displaystyle -\frac{(x_1-z_1)^2}{8a^2}} + H_{12}e^{\displaystyle -\frac{(x_1-x_2)^2}{8a^2}}}z_1 + \frac{H_{12}e^{\displaystyle -\frac{(x_1-x_2)^2}{8a^2}}}{ I_{01}e^{\displaystyle -\frac{(x_1-z_1)^2}{8a^2}} + H_{12}e^{\displaystyle -\frac{(x_1-x_2)^2}{8a^2}}}x_2.
\label{eqA7}
\end{equation}

However, due to the exponential terms in the expression, this result is not compatible with the Gaussian nature of the priors and likelihoods in the Bayesian framework. Rather, it corresponds to an approximately Gaussian likelihood with a slowly broadening function of $x_1-s_1$, namely,
\begin{equation}
p(x_1|s_1) \propto e^{\displaystyle -\frac{(x_1-s_1)^2}{2\sigma(s_1, x_1)^2}} \quad {\rm where} \quad \sigma(s_1, x_1)^2 = \sigma_0^2 e^{\displaystyle -\frac{(x_1-s_1)^2}{8a^2}}. 
\label{eqA8}
\end{equation}
\noindent
This correspondence between the Bayesian framework and the network function is valid in the limit of large $a$. We will not pursue this modification further in this paper.

In Section \ref{Sec3}.\ref{SecC} we consider the regime that the disparity between the two cues is not excessively large compared with the synaptic range $a$ of the networks, wherein the exponential factors in Eq. (\ref{eqA6}) can be neglected, yielding the linear relation Eq. (\ref{neweq21}) in the main text, 
\begin{equation}
x_1 = \frac{I_{01}}{I_{01}+H_{12}}z_1 + \frac{H_{12}}{I_{01}+H_{12}}x_2,
\label{eqA9}
\end{equation}
\noindent
where $H_{ij} = \omega_{ij}R_{0j}/\sqrt{2}$ is the inter-modular contribution from module 2 to module 1 at the maximum position (corresponding to the third term in the right-hand side of Eq. (\ref{eq6}). 

In Section \ref{Sec3}.\ref{SecB} we consider the limit of low disparity and weak inter-modular couplings between modules 1 and 2. Using an isolated module 1 as the reference, we consider the shift of bump position in module 1 when the inter-modular couplings are added.

From Eq. (\ref{eqA3}), the steady-state equation of the bump in an isolated module 1 is given by
\begin{equation}
U_{01} = H_{11} + I_{01},
\label{eqA10}
\end{equation}
\noindent
and the center-of-mass of the bump is located at $z_1$.

In the presence of a weak inter-modular coupling from module 2 to 1, we project the steady-state equation Eq. (\ref{eqA3}) to the displacement mode of the isolated module 1, leading to
\begin{equation}
U_{01}(x_1-z_1) = H_{11}(x_1-z_1) + H_{12}(x_2-z_1).
\label{eqA11}
\end{equation}
\noindent
Noting that the deviation of $U_{01}$ from the solution of Eq. (\ref{eqA10}) belongs to a higher order, we combine Eq. (\ref{eqA9}) with the corresponding equation for module 2, and arrive at
\begin{equation}
\begin{gathered}
\begin{bmatrix}
I_{01} & -H_{12} \\ -H_{21} & I_{02}  
\end{bmatrix}
\begin{bmatrix}
x_1 \\ x_2
\end{bmatrix}
=
\begin{bmatrix}
(I_{01}-H_{12})z_1 \\ (I_{02}-H_{21})z_2.
\end{bmatrix}
\end{gathered}
\label{eqA12}
\end{equation}
\noindent
Therefore, as displayed in Eq. (\ref{neweq12}) in the main text, the displacement of the bump relative to the external input is
\begin{equation}
x_1-z_1 = \frac{H_{12}(I_{02}-H_{21})(z_2-z_1)}{I_{01}I_{02}-H_{12}H_{21}}.
\label{eqA13}
\end{equation}
\noindent
To calculate the change in bump height due to the inter-modular coupling, we project the steady-state equation to the height mode (that is, multiplying both sides of the equation by $e^{(-(x-z_1)^2/4a^2)}$ and integrating $x$). The result is
\begin{equation}
U_{01}e^{-\displaystyle \frac{(x_1-z_1)^2}{8a^2}} = H_{11}e^{-\displaystyle \frac{(x_1-z_1)^2}{8a^2}} + H_{12}e^{-\displaystyle \frac{(x_2-z_1)^2}{8a^2}}+I_{01}.
\label{eqA14}
\end{equation}
\noindent
Comparing with the uncoupled steady-state solution in Eq. (\ref{eqA3}), the change in bump height is given by
\begin{equation}
\delta U_{01}-\frac{(x_1-z_1)^2}{8a^2}U_{01}=\frac{2H_{11}}{B_1U_{01}}\delta U_{01}-\frac{(x_1-z_1)^2}{8a^2}H_{11} + H_{12}e^{-\displaystyle \frac{(x_2-z_1)^2}{8a^2}}.
\label{eqA15}
\end{equation}
\noindent
Simplifying the equation, we obtain
\begin{equation}
\delta U_{01} = \left( 1- \frac{2H_{11}}{B_1U_{01}} \right)^{-1} \left[ H_{12} e^{\displaystyle \frac{(x_2-z_1)^2}{8a^2}} + \frac{(x_1-z_1)^2}{8a^2}I_{01} \right].
\label{eqA16}
\end{equation}
\noindent
The first term contributes to the excitation effect, whereas the displacement effect appears in $(x_1-z_1)^2$ of the second term and belongs to a higher order.

\twocolumngrid
\bibliographystyle{unsrt}
\bibliography{mylib}

\end{document}